\documentclass[reqno]{amsart}

\usepackage{booktabs} % For formal tables
\usepackage{IEEEtrantools}
\usepackage{amssymb,latexsym,amsfonts,amsmath}
\usepackage{graphicx}
\usepackage{mathrsfs}
\usepackage{dsfont}

\usepackage{tikz}
\usetikzlibrary{calc,shapes,arrows}

\usepackage{algorithmic}

\usepackage{xspace}

\usepackage{booktabs} % For formal tables
\usepackage{latexsym,amsfonts,amsmath}
\usepackage{hyperref}
\usepackage{subfigure}
\usepackage{dsfont}
\usepackage{extarrows}
\usepackage[ruled,vlined,linesnumbered]{algorithm2e}

\usepackage{paralist}
\usepackage{capt-of}
\usepackage{xcolor}

\usepackage[many]{tcolorbox}
\usetikzlibrary{calc}
\tcbuselibrary{skins}
\newtcolorbox{resp}[1][]{%
	enhanced jigsaw,%
	colback=gray!5!white,%
	colframe=gray!80!black,%
	size=small,%
	boxrule=1pt,%
	halign title=flush center,%
	coltitle=black,%
	breakable,%
	drop shadow=black!50!white,%
	attach boxed title to top left={xshift=1cm,yshift=-\tcboxedtitleheight/2,yshifttext=-\tcboxedtitleheight/2},%
	minipage boxed title=3cm,%
	boxed title style={%
		colback=white,%
		size=fbox,%
		boxrule=1pt,%
		boxsep=2pt,%
		underlay={%
			\coordinate (dotA) at ($(interior.west) + (-0.5pt,0)$);
			\coordinate (dotB) at ($(interior.east) + (0.5pt,0)$);
			\begin{scope}[gray!80!black]
				\fill (dotA) circle (2pt);
				\fill (dotB) circle (2pt);
			\end{scope}
		}%
	},%
	#1%
}

\newcommand{\N}{{\mathbb{N}}}

\newcommand{\argmax}{\textrm{arg}\max}

\newcommand{\ul}{\underline}

\definecolor{myco}{rgb}{0.55, 0.0, 0.63}

\newcommand\blfootnote[1]{%
	\begingroup
	\renewcommand\thefootnote{}\footnote{#1}%
	\addtocounter{footnote}{-1}%
	\endgroup
}

\newtheorem{theorem}{Theorem}[section]
\newtheorem{lemma}[theorem]{Lemma}
\newtheorem{problem}[theorem]{Problem}
\newtheorem{proposition}[theorem]{Proposition}

\newtheorem{definition}[theorem]{Definition}

\newtheorem{remark}[theorem]{Remark}

\numberwithin{equation}{section}

\begin{document}
	
\begin{abstract}
In this paper, we propose a construction scheme for a Safe-visor architecture for sandboxing unverified controllers, e.g., artificial intelligence-based (a.k.a. AI-based) controllers, in two-players non-cooperative stochastic games. 
Concretely, we leverage abstraction-based approaches to construct a supervisor that checks and decides whether or not to accept the inputs provided by the unverified controller, and a safety advisor that provides fallback control inputs to ensure safety whenever the unverified controller is rejected. 
Moreover, by leveraging an ($\epsilon,\delta$)-approximate probabilistic relation between the original game and its finite abstraction, we provide a formal safety guarantee with respect to safety specifications modeled by deterministic finite automata (DFA), % for any arbitrary adversarial players. 
%At the same time, 
while the functionality of the unverified controllers is still exploited. 
To show the effectiveness of the proposed results, we apply them to a control problem of a quadrotor tracking a moving ground vehicle, in which an AI-based unverified controller is employed to control the quadrotor.
\end{abstract}

\title[Sandboxing (AI-based) Unverified Controllers in Stochastic Games]{Sandboxing (AI-based) Unverified Controllers in Stochastic Games: An Abstraction-based Approach with Safe-visor Architecture}

\author{Bingzhuo Zhong$^{1*}$}
\author{Hongpeng Cao$^{1}$}
\author{Majid Zamani$^{2,3}$}
\author{Marco Caccamo$^{1}$}
\blfootnote{*Corresponding Author.}
\address{$^1$TUM School of Engineering and Design, Technical University of Munich, Germany.}
\email{bingzhuo.zhong@tum.de}
\email{cao.hongpeng@tum.de}
\email{mcaccamo@tum.de}
\address{$^2$Department of Computer Science, University of Colorado Boulder, USA.}
\email{majid.zamani@colorado.edu}
\address{$^3$Department of Computer Science, LMU Munich, Germany.}
\maketitle

\section{Introduction}\label{sec:introd}
{\bf Motivations.} 
The past few decades have witnessed increasing demands in employing unverified (but potentially high performance) controllers, such as artificial intelligence-based (\emph{a.k.a.} AI-based) controllers and black-box controllers from third parties, in cyber-physical systems (CPS)~\cite{Kim2012Cyber} to accomplish complex control missions (e.g.,~\cite{Julian2019Deep}).
However, ensuring the overall safety of the CPS equipped with these controllers is currently very challenging~\cite{Kwiatkowska2019Safety}. 
Meanwhile, safety guarantees are of vital importance for  safety-critical CPS, such as autonomous cars and unmanned aerial vehicles, in which system's failures (\emph{e.g.,} collision) may have catastrophic results.
Hence, the difficulties in providing safety guarantees motivated researchers to investigate correct-by-construction techniques using which formal safety guarantees can be offered.

{\bf Related Works.}
For systems with discrete state and input sets, shields are proposed in~\cite{Humphrey2016Synthesis,Alshiekh2018Safe} to correct erroneous control inputs and enforce safety properties at run-time.
As for systems with continuous state sets, Simplex architecture was introduced in~\cite{Sha2001Using}, which allowed the application of unverified controllers in linear systems by leveraging a Lyapunov-function-based elliptic recovery region (\emph{a.k.a.} safety invariant set). 
This architecture was further improved and extended by allowing more complex dynamics (see e.g.,~\cite{Wang2013L1Simplex,Abdi2018Preserving}), and reducing the conservatism of the recovery regions (see e.g.~\cite{Bak2014Real,Abdi2017Application}).
However, these results can only be applied to enforce invariance properties over deterministic systems (i.e., systems are desired to stay within a set). 
It is also worth mentioning that, when focusing on invariance properties and a particular type of unverified controllers, some recent results (e.g.~\cite{Cheng2019End,Huh2020Safe,Hewing2020Learning}) also provide formal safety guarantees by designing proper reward functions to navigate the training of unverified controllers.
\begin{figure}
	\centering
	\includegraphics[width=8cm]{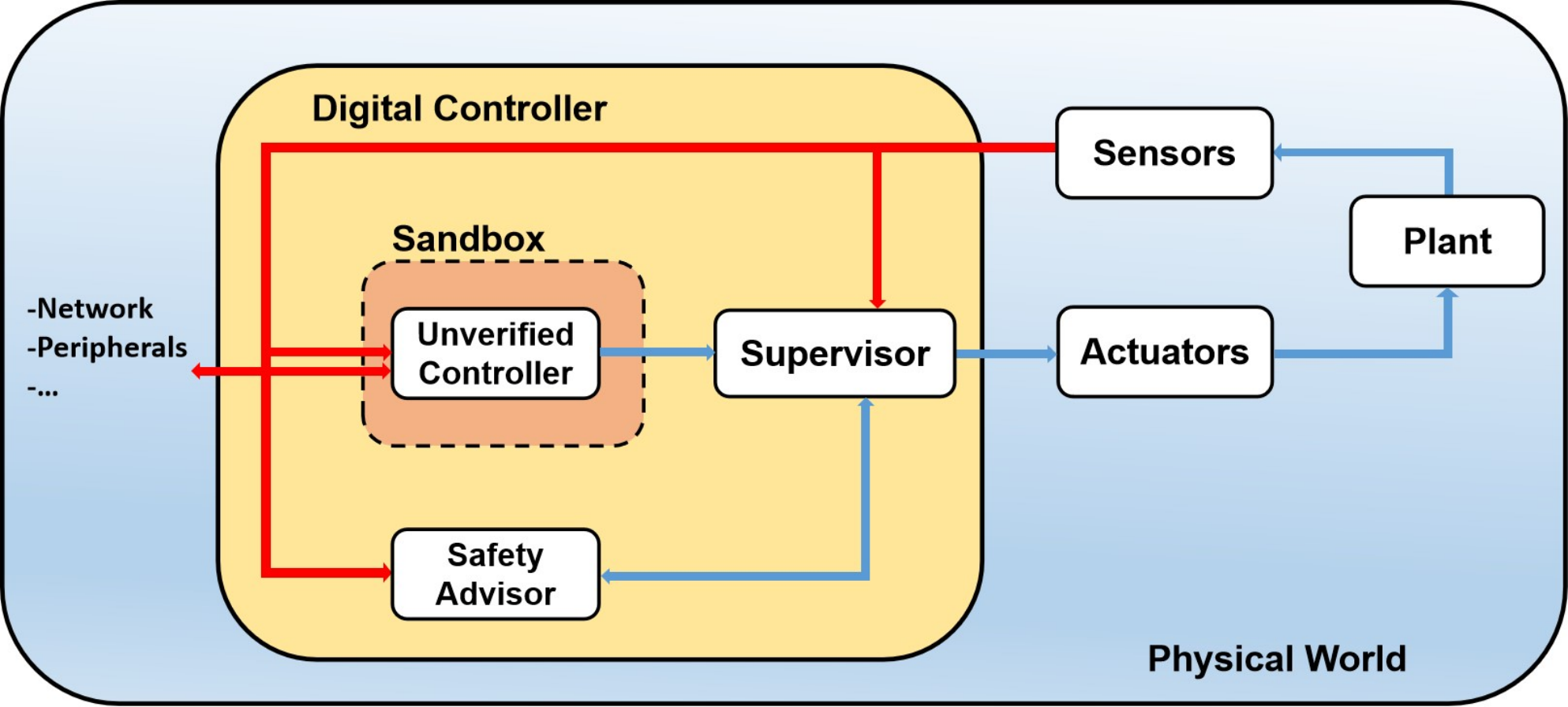}
	\caption{Safe-visor architecture for sandboxing unverified controllers.
		To apply this architecture, one specifies the maximal tolerable probability of violating the desired safety property, denoted by $\eta$.
		At run-time, the supervisor checks the inputs offered by the unverified controller and only accepts them when $\eta$ is respected.
		Whenever the unverified controller is rejected, the safety advisor provides an advisory control command, which maximizes the probability of satisfying the desired safety property, to actuate the plant.} \label{fig1:Svisor_arc}
\end{figure}

When complex logical properties are of interest, \emph{e.g.,} those expressed as linear temporal logic formulae over finite traces (\emph{a.k.a.} LTL$_F$ formulae~\cite{DeGiacomo2013Linear}), results in~\cite{Li2019Temporal,Lavaei2020Formal,Kazemi2020Formal} provide formal safety guarantees for AI-based controllers by considering the desired properties in the reward functions.
Note that these results are only applicable to those AI-based controllers whose reward functions are easy to be designed, while reward functions for some control tasks are difficult to be obtained (e.g.,~\cite{Biyik2022Learning}).
Recently, results in~\cite{Zhong2019Sandboxing} initialized a system level~\cite{Claviere2021Safety}, correct-by-construction architecture, namely \emph{Safe-visor Architecture}, utilizing the idea of \emph{sandbox} that is borrowed from the computer security community~\cite{Reis2009Browser}.
This architecture, as depicted in Figure~\ref{fig1:Svisor_arc}, can be leveraged to enforce invariance properties for continuous-space stochastic systems equipped with arbitrary types of unverified controllers.
Later, ~\cite{Zhong2021Safe} extended the results in~\cite{Zhong2019Sandboxing} by reducing the conservatism of the probabilistic safety guarantee and allowing a larger class of safety specifications modeled by deterministic finite automata (DFA)~\cite{Baier2008Principles}.
Note that in this architecture, the unverified controller is designed for tasks that are more complex than only maintaining the overall safety of the systems.
Hence, the safety advisor is only applied when the unverified controller tries to endanger the safety of the system so that the functionalities offered by the unverified controller can still be exploited.

{\bf Contributions}
Current results for sandboxing controllers, as mentioned above, are only applicable to contin-uous-space systems affected by control inputs (and noise). 
However, in some safety-critical real-life applications, systems are also affected by (rational) adversarial inputs with objectives that are opposed to those of the control inputs.
These systems can be modeled by general discrete-time stochastic games (gDTSG) with two non-cooperative players. 
As the main contribution, we propose a construction scheme for the Safe-visor architecture to sandbox controllers in gDTSGs with respect to safety specifications modeled by DFAs.
Here, we consider stochastic games since they provide a powerful framework to describe many systems operating in noisy environments. 
Additionally, to show the effectiveness of our results, for the first time, we apply the proposed abstraction-based Safe-visor architecture to a physical test-bed (c.f. Section~\ref{Case_study}).

\section{Problem Formulation}\label{sec:2}
\subsection{Preliminaries}
A topological space $S$ is called a Borel space if it is homeomorphic to a Borel subset of a Polish space (\emph{i.e.,} a separable and completely metrizable space). 
Here, any Borel space $S$ is assumed to be endowed with a Borel $\sigma$-algebra, denoted by $\mathcal{B}(S)$. 
A map $f:X\!\rightarrow\! Y$ is measurable whenever it is Borel measurable.

A probability space in this work is presented by $(\hat \Omega,\mathcal F_{\hat \Omega},\mathbb{P}_{\hat \Omega})$, where $\hat \Omega$ is the sample space,
$\mathcal F_{\hat \Omega}$ is a sigma-algebra on $\hat \Omega$ which comprises subsets of $\hat\Omega$ as events, and $\mathbb{P}_{\hat \Omega}$ is a probability measure that assigns probabilities to events.
Throughout the paper, we focus on random variables, denoted by $X$, that take values from measurable spaces $(S,\mathcal{B}(S))$, \emph{i.e.,} random variables here are measurable functions $X:(\hat \Omega,\mathcal F_{\hat \Omega})\rightarrow(S,\mathcal{B}(S))$ such that one has $Prob\{\mathcal{Q}\} = \mathbb{P}_{\hat \Omega}\{X^{-1}(\mathcal{Q})\}$, $\forall \mathcal{Q}\in \mathcal{B}(S)$. 
For succinctness, we directly present the probability measure on $(S,\mathcal{B}(S))$ without explicitly mentioning the underlying probability space and the function $X$ itself. 
Additionally, we denote by $\mathbf{P}(S,\mathcal{B}(S))$ a set of probability measures on the $(S,\mathcal{B}(S))$. 

\subsection{Notations}
We use $\mathbb{R}$ and $\mathbb{N}$ to denote sets of real and natural numbers, respectively. 
These symbols are annotated with subscripts to restrict the sets in a usual way, \emph{e.g.,} $\mathbb{R}_{\geq0}$ denotes the set of non-negative real numbers. 
In addition, $\mathbb{R}^{n\times m}$ with $n,m\in \mathbb{N}_{\geq 1}$ denotes the vector space of real matrices with $n$ rows and $m$ columns.
Given $M\in \mathbb{R}^{n\times m}$, $M^{\top}$ denotes the transpose of $M$.
For $a,b\in\mathbb{R}$ (resp. $a,b\in\mathbb{N}$) with $a\leq b$, the closed, open and half-open intervals in $\mathbb{R}$ (resp. $\mathbb{N}$) are denoted by $[a,b]$, $(a,b)$ ,$[a,b)$ and $(a,b]$, respectively. 
Given $N$ vectors $x_i \in \mathbb R^{n_i}$, $n_i\in \mathbb N_{\ge 1}$, and $i\in\{1,\ldots,N\}$, $x = [x_1;\ldots;x_N]$ denotes the corresponding column vector of dimension $\sum_i n_i$.
Moreover, $\lVert x\rVert$ denotes Euclidean norm of $x$. 
Given a set $X$,  $X^{\mathbb{N}}$ denotes the Cartesian product among the countable infinite number of set $X$.
Furthermore, given sets $X$ and $Y$, a relation $\mathscr{R} \in X\times Y$ is a subset of the Cartesian product $X\times Y$ that relates $x\in X$ with $y\in Y$ if $(x,y)\in\mathscr{R}$, which is equivalently denoted by $x\mathscr{R}y$.
Given functions $f:X\rightarrow Y$ and $g:Y\rightarrow Z$, we denote by $g\circ f : X\rightarrow Z$ the composite function of $f$ and $g$.

\subsection{General Discrete-time Stochastic Games} \label{sec:sys_model}
In this work, we focus on systems that can be formulated as general discrete-time stochastic games (gDTSG), in which control input and adversary input are referred to as Player~\uppercase\expandafter{\romannumeral1} and Player~\uppercase\expandafter{\romannumeral2}, respectively, following standard conventions.
\begin{definition}
	\label{def:gDTSG}
	A gDTSG is a tuple
	\begin{equation}
	\label{eq:dt-SCS}
	\mathfrak{D} =(X,U,W,X_0,T,Y,h),
	\end{equation}
	where
	\begin{itemize}
		\setlength{\itemsep}{0pt}
		\setlength{\parsep}{0pt}
		\setlength{\parskip}{0pt}
		\item $X\subseteq \mathbb R^s$ is a Borel set as the state set. We denote by $(X, \mathcal B (X))$ the measurable space with $\mathcal B (X)$  being  the Borel sigma-algebra on $X$;
		\item $U\subseteq \mathbb R^m$ is a compact Borel set as Player~\uppercase\expandafter{\romannumeral1}'s input set;
		\item $W\subseteq \mathbb R^p$ is a compact Borel set as Player~\uppercase\expandafter{\romannumeral2}'s input set;
		\item $X_0\subseteq X$ is the set of initial states;
		\item $T:\mathcal B(X)\!\times\! X\!\times\! U\!\times\! W\!\rightarrow\![0,1]$ 
		is a conditional stochastic kernel that assigns to any $x \in X$, $u\in U$, and $w\in W$ a probability measure $T(\cdot | x,u,w)$ on $(X,\mathcal B(X))$, such that for any set $\mathcal{Q} \subseteq \mathcal B(X)$ and for any $k\in\mathbb N$, 
		\begin{align*}
		\mathbb P \big\{x(k+1)\in \mathcal{Q}\,\big|\,& x(k),u(k),w(k)\big\}=\int_\mathcal{Q} T (\mathsf dx(k+1)|x(k),u(k),w(k));
		\end{align*}
		\item $Y\subseteq \mathbb R^q$ is a Borel set as the output set; 
		\item $h\!:\!X\!\rightarrow\! Y$ is a measurable output function as $y = h(x)$.
	\end{itemize}
\end{definition}
\begin{remark}\label{rem:asy_info}
	To provide formal probabilistic guarantees regardless of how adversarial inputs are chosen by Player~\uppercase\expandafter{\romannumeral2}, we consider an asymmetric information pattern that favors Player~\uppercase\expandafter{\romannumeral2}.
	Based on this setting, we design the Safe-visor architecture over Player~\uppercase\expandafter{\romannumeral1} considering that Player~\uppercase\expandafter{\romannumeral2} may select its action in a rational fashion against the choice of Player~\uppercase\expandafter{\romannumeral1}.		 
	Note that our setting here is common for robust control problems, while Player~\uppercase\expandafter{\romannumeral2} does not have to select adversarial inputs rationally in a worst-case manner. 
\end{remark}
\begin{remark}
	In this paper, the full state information of the gDTSG is available (rather than only the output), and the properties of interest are described over the output.
\end{remark}

Alternatively, a gDTSG $\mathfrak{D}$ as in~\eqref{eq:dt-SCS} can be described by 
the following difference equations:
\begin{equation}\label{eq:gMDP_f}
\mathfrak{D}\!:
\left\{\hspace{-0.15cm}\begin{array}{l}
x(k+1)=f(x(k),u(k),w(k),\varsigma(k)),\\
y(k)=h(x(k)),\quad \quad k\in\mathbb N,\end{array}\right.
\end{equation}
where $x(k)\in X$, $u(k)\in U$, $w(k)\in W$, $y(k)\in Y$, and $\varsigma:=\{\varsigma(k): \hat\Omega\rightarrow V_{\varsigma}, k\in \N\}$ is a sequence of independent and identically distributed (i.i.d.) random variables from the sample space $\hat\Omega$ to a set $V_{\varsigma}$. 
With this notion, one can leverage the path and output sequences to describe the evolution of a gDTSG, which is defined next.
\begin{definition}\label{def:path}
	A \emph{path} of a gDTSG $\mathfrak{D}$ as in~\eqref{eq:dt-SCS} is 
	\begin{align*}
	\omega\!=\!(x(0),u(0),w(0),\ldots,x(k),u(k),w(k),\ldots),\quad \quad \quad  k\!\in\!\N,
	\end{align*}
	with $x(k)\!\in\!X$, $u(k)\!\in\!U$, and $w(k)\!\in\! W$. 
	We denote by $y_{\omega} $ $= (y(0),y(1),\ldots,y(k),\ldots)$	the \emph{output sequence} associated with $\omega$, with $y(k)=h(x(k))$.
	Moreover, $\omega_k$ and $y_{\omega k}$ denote the path up to time instant $k$ and the corresponding output sequence, respectively.
	The space for all infinite paths $\Omega = (X\times U\times W)^{\N}$ along with its product $\sigma$-algebra $(\mathcal B(X)\times \mathcal B(U)\times \mathcal B(W))^{\N}$ is called a \emph{canonical sample space} for $\mathfrak{D}$.
\end{definition}
Next, we proceed with formulating the safety specifications and the main problem we aim at solving in this work.

\subsection{Safety Specifications and Problem Formulation}\label{sec:DFA}
Here, deterministic finite automata (DFA) is leveraged to model the desired safety specifications, as defined below.
\begin{definition} 
	A deterministic finite automata is a tuple $\mathcal{A}\ =\!(Q, q_0, \Pi,\tau, F)$, where $Q$ is a finite set of states, $q_0\!\in\! Q$ is the initial state, $\Pi$ is a finite set of alphabet, $\tau : Q\times\Pi \rightarrow Q$ is a transition function, and $F\!\subseteq\! Q$ is a set of accepting states.
\end{definition}

Without loss of generality~\cite[Section 4.1]{Baier2008Principles}, we focus on those DFAs which are \emph{total}, i.e., given any $q\in Q$, $\forall \sigma' \in \Pi$, $\exists q'\in Q$ such that $q'\! =\! \tau(q,\sigma')$.
A finite word $\sigma = (\sigma_0, \sigma_1,\ldots,\sigma_{k-1})\in \Pi^k$ is accepted by $\mathcal{A}$ if there exists a finite state run $q\! =\!(q_0,q_1,\ldots,q_k)\!\in\!Q^{k+1}$ such that $q_{z+1}\! =\! \tau(q_z,\sigma_z)$, $\sigma_z\! \in\! \Pi$ for all $0\!\leq\! z\!<\!k$, and $q_k\!\in\! F$.
The set of words accepted by $\mathcal{A}$ is called the language of $\mathcal{A}$ and denoted by $\mathcal{L}(\mathcal{A})$.
Next, we define a measurable labelling function which connects a gDTSG $\mathfrak{D}$ to a DFA $\mathcal{A}$.

\begin{definition}\label{def:sactisfaction_DFA} \emph{(Labelling Function)}
	Consider a gDTSG $\mathfrak{D}\!=\!(X,U,W,X_0,T,Y,h)$, a DFA $\mathcal{A}\!=\! (Q, q_0, \Pi,\tau,$ $ F)$, and a finite output sequence $y_{\omega (H-1)}\!=\!(y(0),y(1),\ldots,y(H-1))\!\in\! Y^H$ of $\mathfrak{D}$ with $H\!\in\!\mathbb{N}_{>0}$.	
	The trace of $y_{\omega (H-1)}$ over $\Pi$ is $\sigma\!=\!L_H(y_{\omega (H-1)})\!=\!$ $(\sigma_0,\sigma_1,\ldots,\sigma_{H-1})$ with $\sigma_k=L(y(k))$ for all $k\in[0,H-1]$, where $L: Y\rightarrow \Pi$ is a measurable labelling function and $L_H:Y^H\rightarrow \Pi^H$ is a measurable function.
	Moreover, $y_{\omega (H-1)}$ is accepted by $\mathcal{A}$, denoted by $y_{\omega (H-1)}\models \mathcal{A}$, if $L_H(y_{\omega (H-1)})\in \mathcal{L}(\mathcal{A})$. 
\end{definition}

Throughout the paper, we denote by $(\mathcal{A},H)$ the specification of interest, with $\mathcal{A}$ being a DFA and $H$ being the finite time horizon over which the specification should be satisfied.
In particular, we focus on those safety specifications for which all infinite output sequences that violate these specifications have a finite bad prefix~\cite[Section 3.3.2]{Baier2008Principles} (\emph{e.g.,} safe-LTL$_F$ properties~\cite{Saha2014Automated}).
Accordingly, we characterize these specifications by DFAs that accept all bad prefixes. Now, we formulate the main problem being tackled here.
\begin{resp}
	\begin{problem}\label{prob:worst_case_violation}
		Consider a gDTSG $\mathfrak{D}$ as in \eqref{eq:dt-SCS}, and a desired specification $(\mathcal{A},H)$.
		Given the maximal tolerable probability of violating $(\mathcal{A},H)$ which is acceptable, denoted by $\eta$ , design a Safe-visor architecture as in Figure~\ref{fig1:Svisor_arc} (if existing) for Player~\uppercase\expandafter{\romannumeral1} of $\mathfrak{D}$ such that
		\begin{align}\label{problem2}
		\mathbb{P}_{\Omega}\Big\{y_{\omega H}\models \mathcal{A}\Big\} \leq \eta
		\end{align}
		holds for any arbitrary adversarial strategies of Player~\uppercase\expandafter{\romannumeral2},
		where $y_{\omega H}$ are output sequences generated by $\mathfrak{D}$.
	\end{problem}
\end{resp}

\section{Design of Safe-visor Architecture} \label{sec:design_sva}
In this section, we first present an approximate probabilistic relation for quantifying the similarity between two different gDTSGs, which plays a crucial role in the Safe-visor architecture for providing formal safety guarantees.
Then, we present the design of Safe-visor architecture in Section~\ref{sec:supervisor}, which is the main result of this paper.
\subsection{Approximate Probabilistic Relations}\label{sec:APR}
First, we present the notation of $\delta$-lifted relation over general state spaces, which is required for defining the approximate probabilistic relation between two gDTSGs.
\begin{definition}\label{lifting}
	\emph{($\delta$-Lifted Relation~\cite{Haesaert2017Verification})}
	Let $X, \hat X$ be two sets with associated measurable spaces $(X, \mathcal B(X))$ and $(\hat X, \mathcal B(\hat X))$.
	Consider a relation	$\mathscr{R}\subseteq X\times \hat X$ that is measurable, i.e., $\mathscr{R} \in \mathcal B(X \times \hat X)$, probability distributions $\Phi\in\mathbf{P}(X, \mathcal B(X))$, and $\Theta\in\mathbf{P}(\hat X, \mathcal B(\hat X))$.
	One has $(\Phi,\Theta)\in\mathscr{\bar R}_{\delta}$, denoted by $\Phi\mathscr{\bar R}_{\delta}\Theta$, with $\mathscr{\bar R}_{\delta}\subseteq  \mathbf{P}(X, \mathcal B(X))\times \mathbf{P}(\hat X, \mathcal B(\hat X))$ being a \emph{$\delta$-lifted relation}, if there exists a probability measure $\mathscr{L}$, referred to as a \emph{lifting}, with a probability space $(X \times \hat X, \mathcal B(X \times \hat{X}), \mathscr{L})$ such that
	\begin{itemize}
		\setlength{\itemsep}{0pt}
		\setlength{\parsep}{0pt}
		\setlength{\parskip}{0pt}
		\item $\forall \mathcal{X} \in \mathcal B(X), ~\mathscr{L}(\mathcal{X} \times \hat X) = \Phi (\mathcal{X})$;
		\item $\forall \mathcal{\hat X} \in \mathcal B(\hat X), ~\mathscr{L}(X \times \mathcal{\hat X}) = \Theta (\mathcal{\hat X})$;
		\item $\mathscr{L}(\mathscr{R})\geq 1-\delta$, i.e., for the probability space $(X \times \hat X, \mathcal B(X \times \hat X), \mathscr{L})$, it holds that $x\mathscr{R} \hat x$ with the probability of at least $1-\delta$.
	\end{itemize}
\end{definition}

Using Definition~\ref{lifting}, we provide conditions under which an ($\epsilon, \delta$)-approximate probabilistic relation~\cite{Haesaert2021Robust} can be established between two gDTSGs. 

\begin{definition}\label{Def: apr} 
	(\emph{($\epsilon,\!\delta$)-Approximate Probabilistic Relation})
	Consider gDTSGs $\mathfrak{D} =(X,U,W,X_0,Y,h)$ and $\widehat{\mathfrak{D}} =(\hat X,\hat U,\hat W,\hat{X}_0,\hat T,Y,\hat h)$.
	The gDTSG $\widehat{\mathfrak{D}}$ is ($\epsilon, \delta $)-stochastically simulated by $\mathfrak{D}$, denoted by $ \widehat{\mathfrak{D}}\preceq_{\epsilon}^{\delta}\mathfrak{D} $, if there exist relations $\mathscr{R}\subseteq X \times \hat X$, $\mathscr{R}_w\subseteq W \times \hat W$ and a Borel measurable stochastic kernel $\mathscr{L}_{T}(\cdot~|~ x, \hat{x},\hat{u}, w, \hat{w})$ on $X \times \hat X$ such that 
	\begin{itemize}
		\setlength{\itemsep}{0pt}
		\setlength{\parsep}{0pt}
		\setlength{\parskip}{0pt}
		\item (Cond. 1) $\forall (x,\hat x)\! \in\! \mathscr{R}$, $\Vert h(x)- \hat h (\hat x) \Vert \!\leq\! \epsilon$;
		\item (Cond. 2) $\forall (x,\hat x) \!\in\! \mathscr{R}$, and $\forall \hat u \!\in\! \hat U$, $\exists u \!\in \!U$ such that $\forall w\!\in\! W$,  $\exists \hat w \in \hat W$ with $(w,\hat w) \in \mathscr{R}_w$ such that one has $T(\cdot| x, u, w)~\mathscr{\bar R}_{\delta} ~ \hat T(\cdot| \hat x,\hat u, \hat w)$ with lifting $\mathscr{L}_{T}(\cdot| x, \hat{x}, \hat{u}, w, \hat{w})$;
		\item (Cond. 3) $\forall x_0\in X_0$, $\exists \hat{x}_0\in\hat{X}_0$ such that $x_0 \mathscr{R}\hat{x}_0 $.
	\end{itemize}
\end{definition}

The second condition of Definition~\ref{Def: apr} implicitly implies that for any $\hat u \in \hat U$, there exists an \emph{interface function} $u:=\nu(x,\hat x, \hat u)\in U$ such that the probability measures of the states of $\mathfrak{D}$ and $\widehat{\mathfrak{D}}$ after one-step transition are in a $\delta$-lifted relation. 
In this paper, we construct a Safe-visor architecture for a gDTSG $\mathfrak{D}$ by first building a \emph{finite abstraction} of $\mathfrak{D}$, denoted by $\widehat{\mathfrak{D}}$, which is a gDTSG with finite state and input sets.
Accordingly, the formal safety guarantee for the desired safety specification is provided based on an ($\epsilon, \delta$)-approximate probabilistic relation between $\mathfrak{D}$ and $\widehat{\mathfrak{D}}$.
To establish such a relation, one can employ existing results (e.g.~\cite[Section 4]{Haesaert2021Robust}, ~\cite[Section 4]{Zhong2021Automata}).
Due to lack of space, we are not providing details of those results.

\subsection{Design of Safe-visor Architecture}\label{sec:supervisor}
\begin{figure}
	\centering
	\includegraphics[width=7cm]{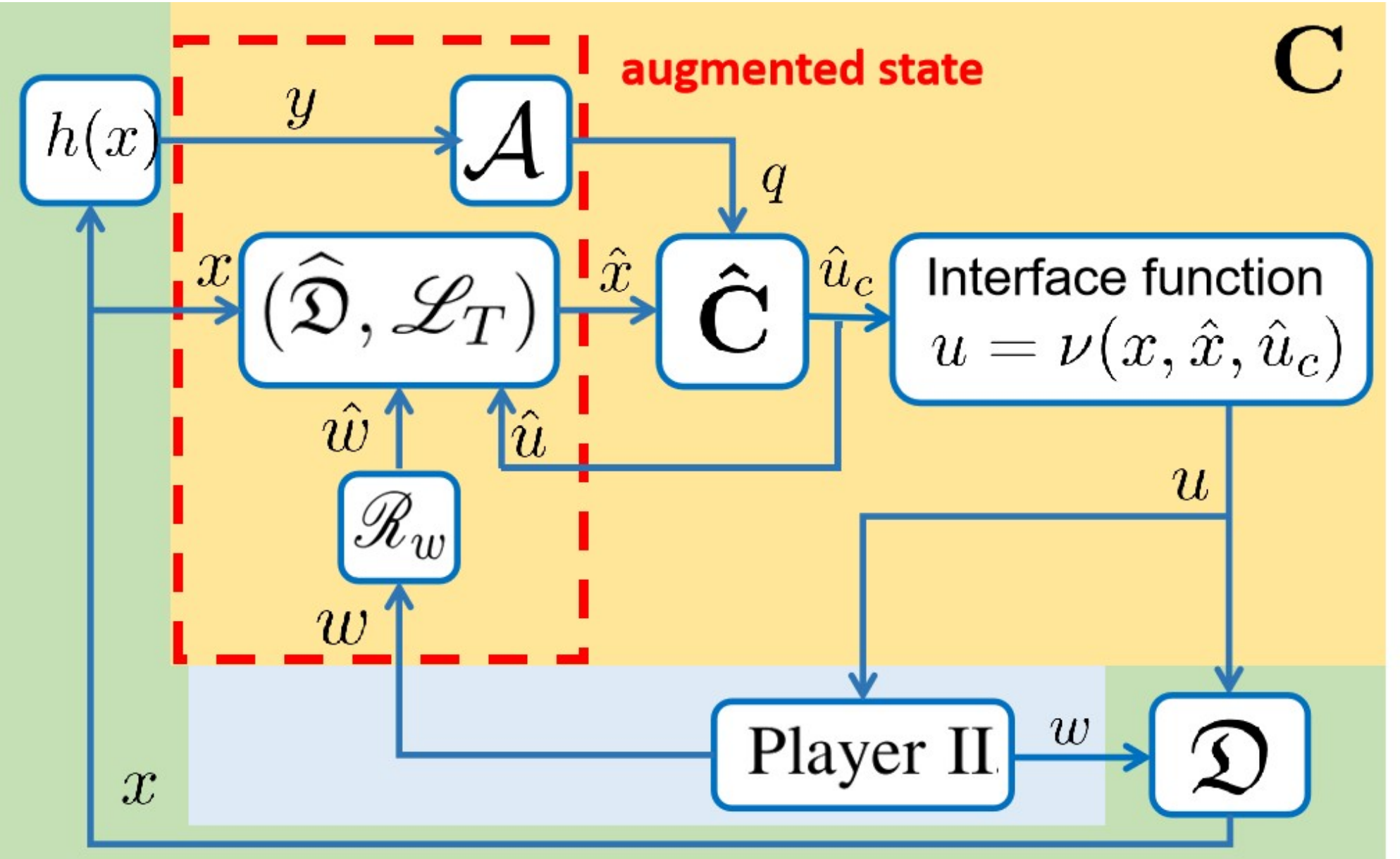}
	\caption{Safety advisor (yellow region) , which is a controller over $\mathfrak{D}$ (green region), with augmented state $(x,\hat{x},q,\hat{u},w)$ (red dashed rectangle).} \label{fig1:sa}
\end{figure}

In general, the construction of a Safe-visor architecture includes offline computation of a safety advisor enforcing the desired safety specification, and online implementation of a supervisor that detects harmful inputs from the unverified controller and decides the input fed to the system accordingly (either the one from the unverified controller or the safety advisor).
Here, we leverage the results in~\cite{Zhong2021Automata} for synthesizing the safety advisor.
Due to space limitations, we omit their technical details here.
Instead, we summarize the general idea for constructing the safety advisor as follows:
\begin{itemize}
	\item Given a gDTSG $\mathfrak{D}$, we first build its finite abstraction $\widehat{\mathfrak{D}}$ using the results in~\cite[Section 4]{Zhong2021Automata} such that $\widehat{\mathfrak{D}}\preceq^{\delta}_{\epsilon}\mathfrak{D}$. 
	In brief, we partition the continuous state and input states of $\mathfrak{D}$ via finite numbers of bounded cells and select representative points for these cells to construct $\widehat{\mathfrak{D}}$.
	The outcome is a matrix $\hat T$ representing the dynamics of $\widehat{\mathfrak{D}}$.
	\item Then, we synthesize a controller, denoted by $\mathbf{\hat{C}}$, over $\widehat{\mathfrak{D}}$ by leveraging~\cite[Proposition 5.11]{Zhong2021Automata}, based on which we construct the safety advisor, denoted by $\mathbf{C}$, according to~\cite[Definition 5.1]{Zhong2021Automata}.
\end{itemize}
The safety advisor $\mathbf{C}$ is depicted in Figure~\ref{fig1:sa}.
Here, the safety advisor utilizes an augmented state $(x,\hat{x},q,\hat{u},w)$, which contains states $x$, $\hat{x}$, and $q$ of $\mathfrak{D}$, $\widehat{\mathfrak{D}}$, and $\mathcal{A}$, respectively, the control input $\hat{u}$ fed to $\widehat{\mathfrak{D}}$, and the adversary input $w$ from the Player~\uppercase\expandafter{\romannumeral2} of $\mathfrak{D}$.
%By leveraging this augmented state, g
Given a state $x$ of $\mathfrak{D}$ at time instant $k$, the running mechanism of $\mathbf{C}$ is summarized in Algorithm~\ref{tbl:update_mem_state}.
\begin{algorithm}
	\caption{Running mechanism of safety advisor.}
	\label{tbl:update_mem_state}
	\KwIn{A gDTSG $\mathfrak{D}$, a safety property $(\mathcal{A},H)$ with $\mathcal{A}=(Q, q_0,$ $ \Pi,\tau, F)$, safety advisor $\mathbf{C}$, 
		and the current state $x(k)$ of $\mathfrak{D}$.}
	\uIf{$k=0$}{
		Update $\hat{x}(k)$ such that  $x(k) \mathscr{R}\hat{x}(k)$ (cf. (\emph{Cond. 3}) of Definition~\ref{Def: apr});
	}
	\Else{
		Update $\hat{x}(k)$ according to $x(k)$, the conditional stochastic kernel $\mathscr{L}_T$, and $w(k-1)$ (cf. (\emph{Cond. 2}) of Definition~\ref{Def: apr});
	}
	Update $q$ of $\mathcal{A}$ as $q(k)=\tau(q(k-1),L\circ h(x(k)))$;\\
	Refine $\hat{u}_c$, which is offered by $\mathbf{\hat{C}}$ based on $\hat{x}(k)$ and $q(k)$, to $\mathfrak{D}$ with the interface function $\nu$ (cf. Figure~\ref{fig1:sa});\\
	Updates $\hat{u}(k)$ as $\hat{u}(k): = \hat{u}_c$;\label{update_uhat}\\
	Update $w(k)$ after Player~\uppercase\expandafter{\romannumeral2} has made decision.\\
	\KwOut{$u(k)$ for controlling the system $\mathfrak{D}$.}
\end{algorithm}
\begin{figure}
	\centering
	\includegraphics[width=7cm]{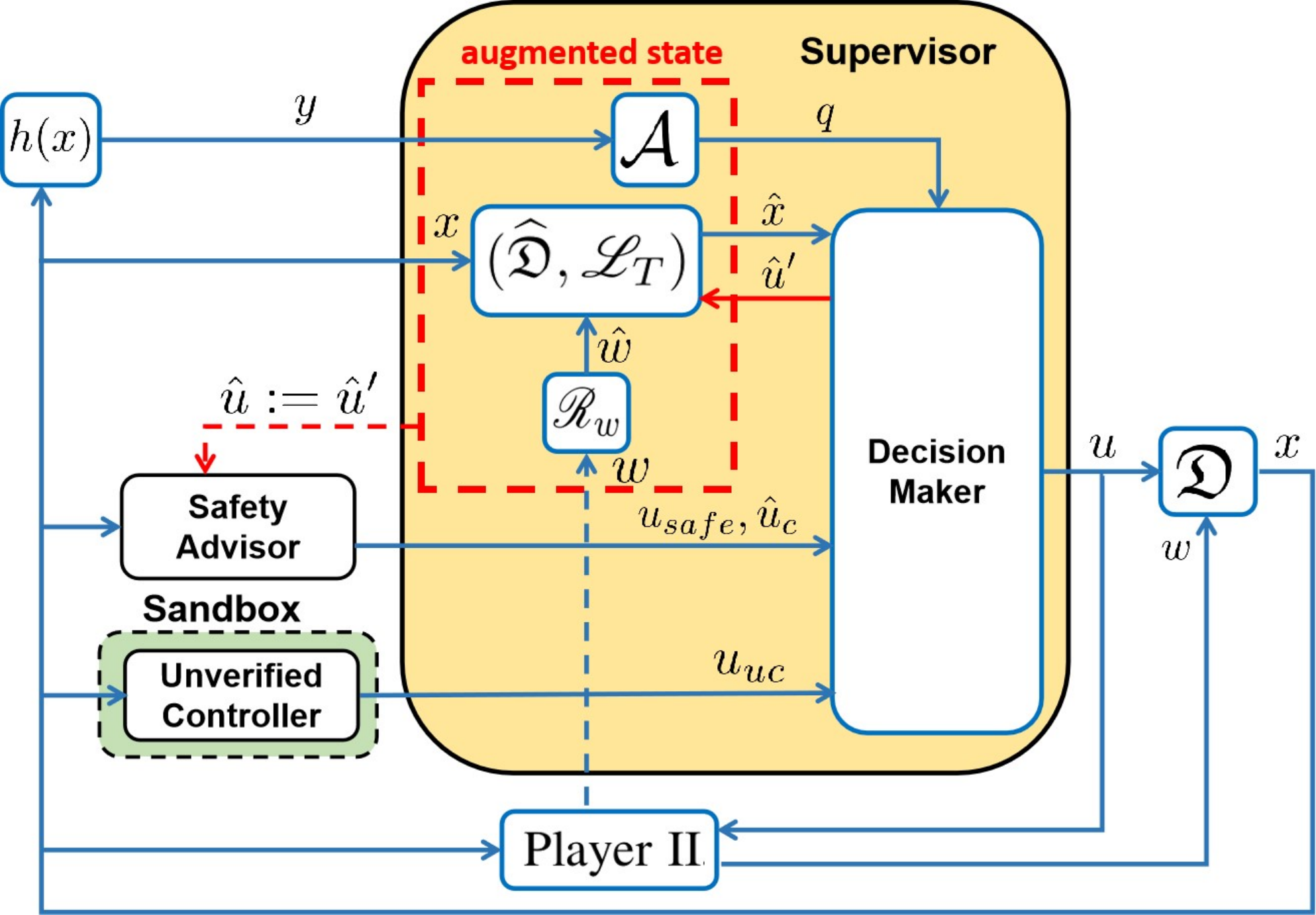}
	\caption{Supervisor in the Safe-visor architecture (yellow region).} \label{fig:supervisor}
\end{figure}
\begin{remark}\label{re:sa_guarantee}
	By refining the controller $\mathbf{\hat{C}}$ over $\widehat{\mathfrak{D}}$ to $\mathfrak{D}$ with the interface function (see Figure~\ref{fig1:sa}), 
	one has $\Vert y- \hat{y} \Vert \!\leq\! \epsilon$ with a probability of at least $1-\delta$ at each time step considering the ($\epsilon,\delta$)-approximate probabilistic relation between $\widehat{\mathfrak{D}}$ and $\mathfrak{D}$.
	Such distance-based relation  plays a crucial role in the Safe-visor architecture to provide safety guarantees.
	As a key insight, $y$ is controlled indirectly by controlling $\hat{y}$ while keeping $y$ sufficiently closed to $\hat{y}$ with some probability.
\end{remark}
\begin{remark}
	By leveraging the results in~\cite[Corollary 5.12]{Zhong2021Automata}, one gets an upper bound on $\mathbb{P}_{\Omega}\{y_{\omega H}\models \mathcal{A}\}$, denoted by $\mathbf{v}$, when the safety advisor is applied over time horizon $[0,H]$.
	Accordingly,~\eqref{problem2} is achievable whenever $\eta\geq \mathbf{v}$.
\end{remark}

Except for the safety advisor $\mathbf{C}$, one also obtain a cost-to-go function $\ul{V}_{*}$ associated with $\mathbf{C}$, which would be used in the supervisor (cf. Definition~\ref{def:History-based Supervisor_wcv}).
The computation of $\ul{V}_{*}$ is presented as in Definition~\ref{Vstar}, which is adapted from~\cite{Zhong2021Automata}.
\begin{definition}\label{Vstar}
	Consider a gDTSG $\mathfrak{D} =(X,U,W,X_0,Y,$ $h)$ and its finite abstraction $\widehat{\mathfrak{D}} =(\hat X,\hat U,\hat W,\hat{X}_0,\hat T,Y,$ $\hat h)$ with $\widehat{\mathfrak{D}}\preceq^{\delta}_{\epsilon}\mathfrak{D}$, and the desired safety property $(\mathcal{A},H)$ with $\mathcal{A} = (Q, q_0, \Pi,\tau, F)$.
	We define a cost-to go function $\ul{V}_{*,n}:\hat{X}\times Q\rightarrow [0,1]$, which is initialized as $\ul{V}_{*,n+1}(\hat{x},q)$ $=1$ when $q\in F$, and $\ul{V}_{*,n+1}(\hat{x},q)=0$, otherwise, and recursively computed as
	\begin{align*}
	&\ul{V}_{*,n+1}(\hat{x},q):=\!\left\{
	\begin{aligned} 
	&\min_{\hat{u}\in\hat{U}}\max_{\hat{w}\in\hat{W}}\big((1-\delta)\!\!\sum_{\hat{x}'\in \hat{X}}\ul{V}_{*,n}(\hat{x}',\overline{q}_*(\hat{x}',q))\hat{T}(\hat{x}'|\hat{x},\hat{u},\hat{w})+\delta\big),\text{if } q\notin F;\\
	& \quad \quad \quad \quad 1, \quad \quad \quad \quad \quad \quad \quad \quad\quad \quad \quad \quad\quad \quad \quad \quad\quad \quad \quad \ \,  \text{if } q\in F,
	\end{aligned}\right.
	\end{align*}
	with $\overline{q}_*(\hat{x}',q) := \mathop{\arg\max}_{q'\in Q'_{\epsilon}(\hat{x}')}\ul{V}_{*,n}(\hat{x}',q')$,
	$Q'_{\epsilon}(\hat{x}') \!:=\! \Big\{q'\!\in\! Q\mid\!\exists x\!\in\! X,q'\!=\! \tau(q,L\!\circ\! h(x)),\text{with}\ h(x)\!\in\! {N}_{\epsilon}(\hat{h}(\hat{x}'))\Big\}$,
	and 
	$\mathcal{N}_{\epsilon}(\hat{y}):=\{y\in Y\,|\, \lVert y-\hat{y}\rVert\leq \epsilon \}$.
\end{definition}

Next, we proceed with discussing the design of supervisor, which is the main contribution of this paper.
Given a gDTSG $\mathfrak{D}$ and a desired safety specification $(\mathcal{A},H)$, the design of supervisor is depicted in Figure~\ref{fig:supervisor}.
Here, the supervisor consists of a \emph{augmented state} and a \emph{decision maker}.
The augmented state of the supervisor, denoted by $(x,\hat{x},q,\hat{u},w)$, is the same as that of the safety advisor, and we simply say the augmented state of the Safe-visor architecture in the rest of this paper for the sake of brevity.
At run-time, $x$, $\hat{x}$, $q$, and $w$ in the augmented states are updated as described in Algorithm~\ref{tbl:update_mem_state}.
Meanwhile, different from step~\ref{update_uhat} of Algorithm~\ref{tbl:update_mem_state}, $\hat{u}$ here is updated as $\hat{u}:=\hat{u}'$, in which $\hat{u}'$ is determined based on the decision of the supervisor (cf. Definition~\ref{def:History-based Supervisor_wcv}).
With the help of the augmented state, the decision maker of the supervisor decides whether or not to accept the input from the unverified controller at time instant $k$, denoted by $u_{uc}(k)$, in the following way:
\begin{itemize}
	\item Step 1: Assume that $u_{uc}(k)$ is accepted.
	If the ($\epsilon$,$\delta$)-approximate probabilistic relation between $\mathfrak{D}$ and $\widehat{\mathfrak{D}}$ does not hold any more, reject $u_{uc}(k)$ without going through Step 2 and feed input from the safety advisor, denoted by $u_{\text{safe}}$, to $\mathfrak{D}$; proceed to Step 2, otherwise; 
	\item Step 2: Estimate the probability of violating $(\mathcal{A},H)$, denoted by $\mathcal{E}_{pv}(k)$, assuming that $u_{uc}(k)$ is accepted.	Accept $u_{uc}(k)$ if $\mathcal{E}_{pv}(k)\!\leq\! \eta$; otherwise, feed $u_{\text{safe}}$ to $\mathfrak{D}$.
\end{itemize}
Step 1 aims at maintaining the ($\epsilon,\delta$)-approximate probabilistic relation between $\mathfrak{D}$ and $\widehat{\mathfrak{D}}$, which is crucial for providing safety guarantee (cf. Remark~\ref{re:sa_guarantee}).
One can check Step 1 with the following proposition.
\begin{resp}
	\begin{proposition}\label{PR_for_UC}
		Consider a gDTSG $\mathfrak{D}\!\!=\!\!(X,U,W,X_0,Y,h)$ and its finite abstraction $\widehat{\mathfrak{D}} =(\hat X,\hat U,\hat W,$ $\hat{X}_0,\hat T,Y,\hat h)$ with $\widehat{\mathfrak{D}}\preceq^{\delta}_{\epsilon}\mathfrak{D}$ with respect to the relation $\mathscr{R}$ and $\mathscr{R}_w$ as in Definition~\ref{Def: apr}.
		If the set $U_f$ in~\eqref{eq:Uf} is not empty, then the ($\epsilon$,$\delta$)-approximate probabilistic relation still holds between $\mathfrak{D}$ and $\widehat{\mathfrak{D}}$ at the time instant $k+1$ when $u_{uc}(k)$ is applied to $\mathfrak{D}$ at the time instant $k$.
	\end{proposition}
\end{resp}
\begin{figure*}
	\begin{small}
		\rule[0pt]{\textwidth}{0.05em}
		\begin{align}
		&
		\!\!\!\!\!\!\!\!U_f\!:=\!\Big\{\hat{u}\in \hat{U}\big|\forall(w,\hat w) \!\in \!\mathscr{R}_w, \mathbb{P}\big\{(x',\hat{x}')\!\in\! \mathscr{R}\big\}\!\geq\! 1\!-\!\delta\text{ holds}, \text{with }x'\!=\! f(x(k),u_{uc}(k),w,\varsigma(k)),\hat{x}'\!=\!\hat{f}(\hat{x}(k),\hat{u},\hat{w},\hat{\varsigma}(k))\Big\},\label{eq:Uf}\\
		&\mathcal{C}_1(k) := \prod_{z=1}^{k}\Big((1-\delta)\min_{\hat{w}\in \hat{W}}\sum_{\hat{x}\in \hat{X}'_{-\epsilon}(q(z-1))}\!\!\!\!\!\!\!\hat{T}\big(\hat{x}\,\big|\,\hat{x}(z-1), \hat{u}(z-1),\hat{w}\big)\Big),\label{eq:C1}\\
		&\mathcal{C}_2(k) := (1-\delta)\Big(1-\max_{\hat{w}\in \hat{W}}\sum_{\hat{x}\in \hat{X}}\ul{V}_{*,H-k-1}\big(\hat{x},\bar{q}_*(\hat{x}(k),q(k))\big)\hat{T}\big(\hat{x}\,\big|\,\hat{x}(k),\hat{u}^*,\hat{w}\big)\Big)\label{eq:C2},
		\end{align}
		\rule[0pt]{\textwidth}{0.05em}
	\end{small}
\end{figure*}
For the set $U_f$ in~\eqref{eq:Uf},
$f$ and $\hat{f}$ are transition maps of $\mathfrak{D}$ and $\widehat{\mathfrak{D}}$, respectively, as in~\eqref{eq:gMDP_f}, and 
$x(k)\in X$, $\hat{x}(k)\in \hat{X}$, $\varsigma(k)$ and $\hat{\varsigma}(k)$ denote current states of $\mathfrak{D}$ and $\widehat{\mathfrak{D}}$, noise affecting $\mathfrak{D}$ and  $\widehat{\mathfrak{D}}$, respectively. 
As a key insight, $U_f\neq \emptyset$ ensuring that there exists at least one $\hat{u}\in \hat{U}$ corresponding to $u_{uc}(k)$ such that $(x(k+1),\hat{x}(k+1))\in \mathscr{R}$ holds with the probability of at least $1-\delta$.
In general, checking the non-emptiness of $U_f$ depends on the concrete form of $f$.
In Section~\ref{Case_study}, we show how to check whether $U_f$ is empty using Proposition~\ref{PR_for_UC} via the case study.
Next, with the help of Proposition~\ref{PR_for_UC}, we present the supervisor for Problem~\ref{prob:worst_case_violation} as follows.

\begin{definition}\label{def:History-based Supervisor_wcv}
	Consider a gDTSG $\mathfrak{D} =(X,U,W,X_0,Y,h)$ and its finite abstraction $\widehat{\mathfrak{D}} =(\hat X,\hat U,\hat W,\hat{X}_0,\hat T,Y,$ $\hat h)$ with $\widehat{\mathfrak{D}}\preceq^{\delta}_{\epsilon}\mathfrak{D}$, the desired safety specification $(\mathcal{A},H)$ with $\mathcal{A} = (Q, q_0, \Pi,\tau, F)$, a labelling function $L: Y\rightarrow \Pi$ associated with $\mathcal{A}$ as in Definition~\ref{def:sactisfaction_DFA}, and $\eta$ to be the maximal tolerable probability of violating $(\mathcal{A},H)$. 
	At each time instant $k\in [0,H-1]$, the validity of an input $u_{uc}(k)$ from the unverified controller is checked as follows:
	\begin{enumerate}[(i)]
		\item Reject $u_{uc}(k)$ if $U_f$ as in~\eqref{eq:Uf} is empty;
		\item If $U_f$ is not empty, compute $\mathcal{E}_{pv}(k)$ as
		\begin{align}
		\mathcal{E}_{pv}(k):=1-\mathcal{C}_1(k)\mathcal{C}_2(k),\label{eq:supv_safe}
		\end{align}
		in which $\mathcal{C}_1(k)$ and $\mathcal{C}_2(k)$ are computed as in~\eqref{eq:C1} and~\eqref{eq:C2}, respectively, $\ul{V}_{*,H-k-1}$ and $\bar{q}_*$ are defined as in Definition~\ref{Vstar},
		\begin{equation}\label{eq:u*2}
		\hat{u}^* := \argmax_{\hat{u}\in U_f}\ul{\mathcal{E}}_{pv}(k),
		\end{equation}
		and 
		$\hat{X}'_{-\epsilon}(q(z-1))\!\!:=\!\Big\{\hat{x}\!\in\!\hat{X}\big|\forall x\in X,\tau(q(z-1),L\!\circ\! h(x))\!\notin\!F, h(x)\!\in\!{N}_{\epsilon}(\hat{h}(\hat{x}))\!\Big\}$,
		with ${N}_{\epsilon}(\hat{h}(\hat{x}))$ being as in Definition~\ref{Vstar}.
		If $\mathcal{E}_{pv}(k)\!\leq\! \eta$, the supervisor accepts $u_{uc}(k)$ and update the augmented state with $\hat{u}' := \hat{u}^*$ (cf. Figure~\ref{fig:supervisor}), where $\hat{u}^*$ is computed as in~\eqref{eq:u*2}; otherwise, it rejects $u_{uc}(k)$ and set $\hat{u}'$ as $\hat{u}' := \hat{u}_c$, with $\hat{u}_c$ provided by the safety advisor (cf. Algorithm~\ref{tbl:update_mem_state}).
	\end{enumerate}
\end{definition}

By leveraging the supervisor in Definition~\ref{def:History-based Supervisor_wcv}, we propose the main result of this paper.
\begin{resp}
	\begin{theorem}\label{theorem:guarantee for wcv}
		Consider a gDTSG $\mathfrak{D} =(X,U,W,X_0,Y,h)$ and a safety specification $(\mathcal{A},H)$. 
		By leveraging the supervisor in Definition~\ref{def:History-based Supervisor_wcv} at all time $k\in[0,H-1]$ in the Safe-visor architecture for Player~\uppercase\expandafter{\romannumeral1} of $\mathfrak{D}$, one has
		\begin{equation}\label{eq:gua_safe}
		\mathbb{P}_{\Omega}\Big\{y_{\omega H}\models \mathcal{A}\Big\}\leq \eta,
		\end{equation}
		holds for any arbitrary adversarial strategies of Player~\uppercase\expandafter{\romannumeral2},
		where $y_{\omega H}$ are output sequences of $\mathfrak{D}$.
	\end{theorem}
\end{resp}
The proof of Theorem~\ref{theorem:guarantee for wcv} is provided in the Appendix. 
The running mechanism of the proposed Safe-visor architecture is summarized in Algorithm~\ref{tbl:safe-visor_mech}. 
Furthermore, we also provide intuition for $\mathcal{C}_1(k)$ and $\mathcal{C}_2(k)$ and discuss their computational complexity as follows:
\begin{itemize}
	\item $\mathcal{C}_1(k)$ denotes the minimal probability of $\mathfrak{D}$ not being accepted by $\mathcal{A}$ over the time horizon $[0,k]$, while one has $x(z)\mathscr{R}\hat{x}(z)$, $\forall z\in[0,k]$.
	Considering the definition of $\hat{X}'_{-\epsilon}(q(z-1))$ as in Definition~\ref{def:History-based Supervisor_wcv}, one can verify $\forall \hat{x}\!\in\!\hat{X}'_{-\epsilon}(q(z-1))$ with $q(z-1)\!\in\! Q$, $\nexists x$ with $x\mathscr{R}\hat{x}$ such that $F$ is reached at time step $z$.
	In other words, if $\hat{x}(z)\in\hat{X}'_{-\epsilon}(q(z-1))$, one can ensure that $\mathfrak{D}$ will not be accepted by $\mathcal{A}$ at the time $z$ by ensuring $x(z)\mathscr{R}\hat{x}(z)$.
	Hence, given $\hat{x}(z-1)$, $\hat{u}(z-1)$, and $q(z-1)$, 
	\begin{align*}
	\mathcal{C}'(z)\!:=\!(1\!-\!\delta)\min_{\hat{w}\in \hat{W}}\!\!\!\sum_{\hat{x}\in \hat{X}'_{-\epsilon}(q(z-1))}\!\!\!\!\!\!\!\!\!\!\!\hat{T}(\hat{x}\big|\hat{x}(z\!-\!1), \hat{u}(z\!-\!1),\hat{w}),
	\end{align*}
	denotes the minimal probability of $\mathfrak{D}$ not being accepted by $\mathcal{A}$ at time $z$ while $x(z)\mathscr{R}\hat{x}(z)$ still holds.
	As for the complexity of computing $\mathcal{C}_1(k)$, one can verify that $\mathcal{C}_1(k)=\mathcal{C}_1(k-1)	\mathcal{C}'(k)$.
	On one hand, since $\mathcal{C}_1(k-1)$ has already been computed at time step $k-1$, one only needs $\mathcal{C}'(k)$ to obtain $\mathcal{C}_1(k)$ at time $k$.
	On the other hand, $\forall q\in Q$, set $\hat{X}'_{-\epsilon}(q)$ can be computed offline, and $\hat{T}$ is readily computed when synthesizing the safety advisor. 
	Therefore, the number of operations required for computing $\mathcal{C}_1(k)$ at time instant $k$ is proportional to the cardinality of the set $\hat{X}'_{-\epsilon}(q(k-1))$ and $\hat{W}$.
	\item $\mathcal{C}_2(k)$ denotes the probability of $\mathfrak{D}$ not being accepted by $\mathcal{A}$ over the time horizon $[k+1,H]$, while (i) $x(k+1)\mathscr{R}\hat{x}(k+1)$ holds, given $\hat{x}(k)$, $q(k)$, and $\hat{u}(k)=\hat{u}^*$; (ii) $\mathfrak{D}$ is controlled by the safety advisor within $[k+1,H]$.
	Since $\ul{V}_{*,H-k-1}$ and $\hat{T}$ are readily computed when synthesizing the safety advisor, the number of operations required for computing $\mathcal{C}_2(k)$ is proportional to the number of elements in sets $\hat{X}$, $\hat{W}$, and $U_f$.
\end{itemize}
In conclusion, the number of operations required for computing $\mathcal{E}_{pv}(k)$ in~\eqref{eq:supv_safe} is proportional to cardinality of the sets $\hat{X}$, $\hat{W}$, and $U_f$.
We show the real-time applicability of the Safe-visor architecture in Section~\ref{Case_study} via the case study.
\vspace{-0.4cm}
\begin{algorithm}
	\caption{Running mechanism of the Safe-visor.}
	\label{tbl:safe-visor_mech}
	\KwIn{A gDTSG $\mathfrak{D}$, a safety specification $(\mathcal{A},H)$, the Safe-visor architecture with safety advisor $\mathbf{C}$, and supervisor as in Definition~\ref{def:History-based Supervisor_wcv}).
	}
	$k=0$, $x(0) = x_0$\\
	\While{$k<H$}
	{
		Compute $\mathcal{C}_1(k)$ as in~\eqref{eq:C1};\\
		Update $x(k)$, $\hat{x}(k)$, and $q(k)$ in the augmented state as in Algorithm~\ref{tbl:update_mem_state} ;\\
		Obtain $u_{uc}(k)$ from the unverified controller;\\
		\uIf{$u_{uc}(k)~\text{is accepted}$}{
			Set $u(k)=u_{uc}(k)$ and update $\hat{u}(k)$ in the augmented state as $\hat{u}(k)=\hat{u}^*$ with~\eqref{eq:u*2};
		}
		\Else{
			Obtain $u_{\text{safe}}(k)$ and $\hat{u}_c(k)$ from the safety advisor, set $u(k)=u_{\text{safe}}(k)$, and update $\hat{u}(k)$ in the augmented state as $\hat{u}(k)=\hat{u}_c(k)$;
		}%\\
		Update $w(k)$ in the augmented state based on the decision of Player~\uppercase\expandafter{\romannumeral2}.\\
		$k=k+1$
	}
	\KwOut{$u(k)$ for controlling $\mathfrak{D}$ at time instant $k$.}
\end{algorithm}

\section{Case Studies}\label{Case_study}
\begin{figure}[htbp]
	\centering
	\subfigure[Quadrotor and vehicle.]{
		\includegraphics[width=4.5cm]{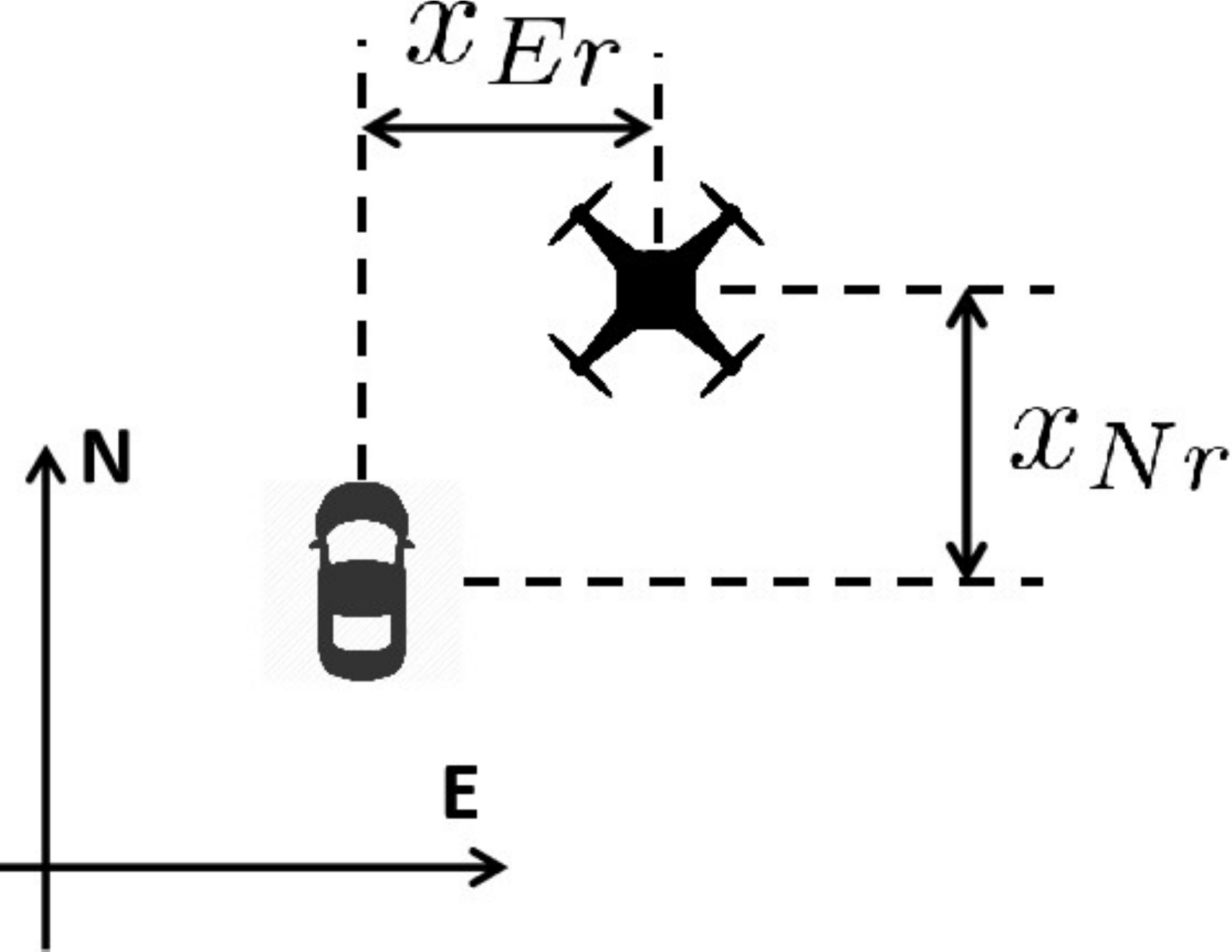}\hspace{-0.4cm}
		%\caption{fig1}
	}
	\quad
	\subfigure[Cross-style quadrotor.]{
		\includegraphics[width=4.5cm]{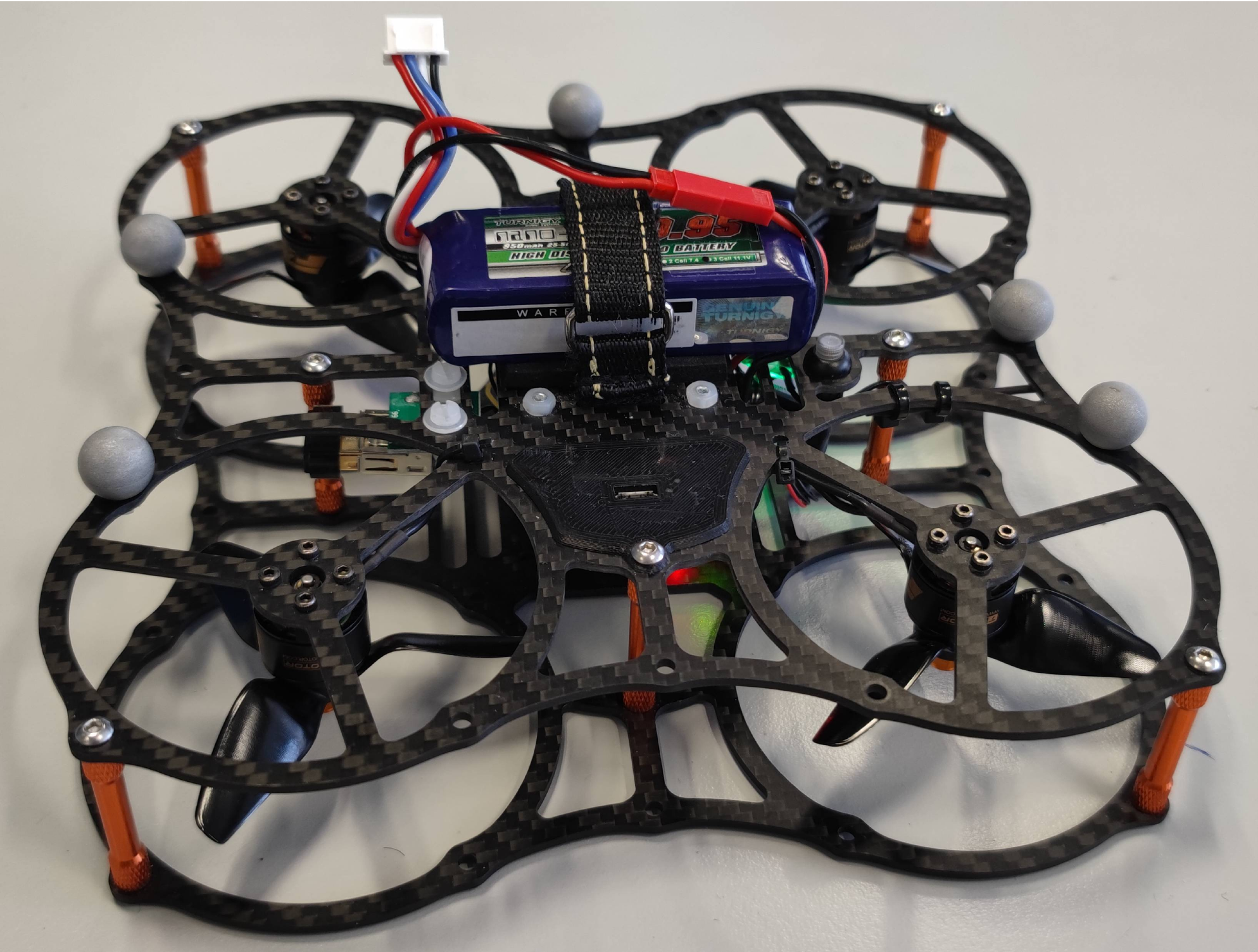}
	}
	\caption{Case study for controlling a quadrotor tracking a ground vehicle.}\label{fig:experiment_setup}
\end{figure}

To show the effectiveness of our proposed results, we apply them to a case study of controlling a quadrotor helicopter tracking a ground vehicle.
Here, we demonstrate the case study through 1) simulation with $1.0\times 10^5$ different realization of noise; 2) experiment on the physical test-bed.
The physical test-bed includes: 
1) a cross-style quadrotor as in Figure~\ref{fig:experiment_setup}; 
2) Vicon motion capture system for capturing the position and velocity of the quadrotor at run-time; and
3) a ground control station (GCS) with Ubuntu 20.04 (Intel Core i9-10900K CPU (3.7 GHz) and 32 GB of RAM).
The simulations are performed via MATLAB 2019b on the GCS.
\begin{figure}[htbp]
	\centering
	\subfigure{
		\includegraphics[width=2cm]{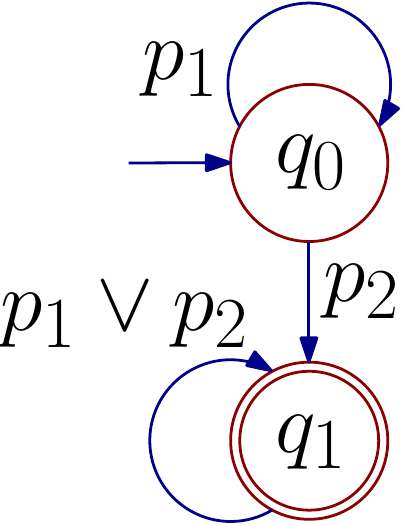}\hspace{-0.4cm}
		%\caption{fig1}
	}
	\quad
	\subfigure{
		\includegraphics[width=7.5cm]{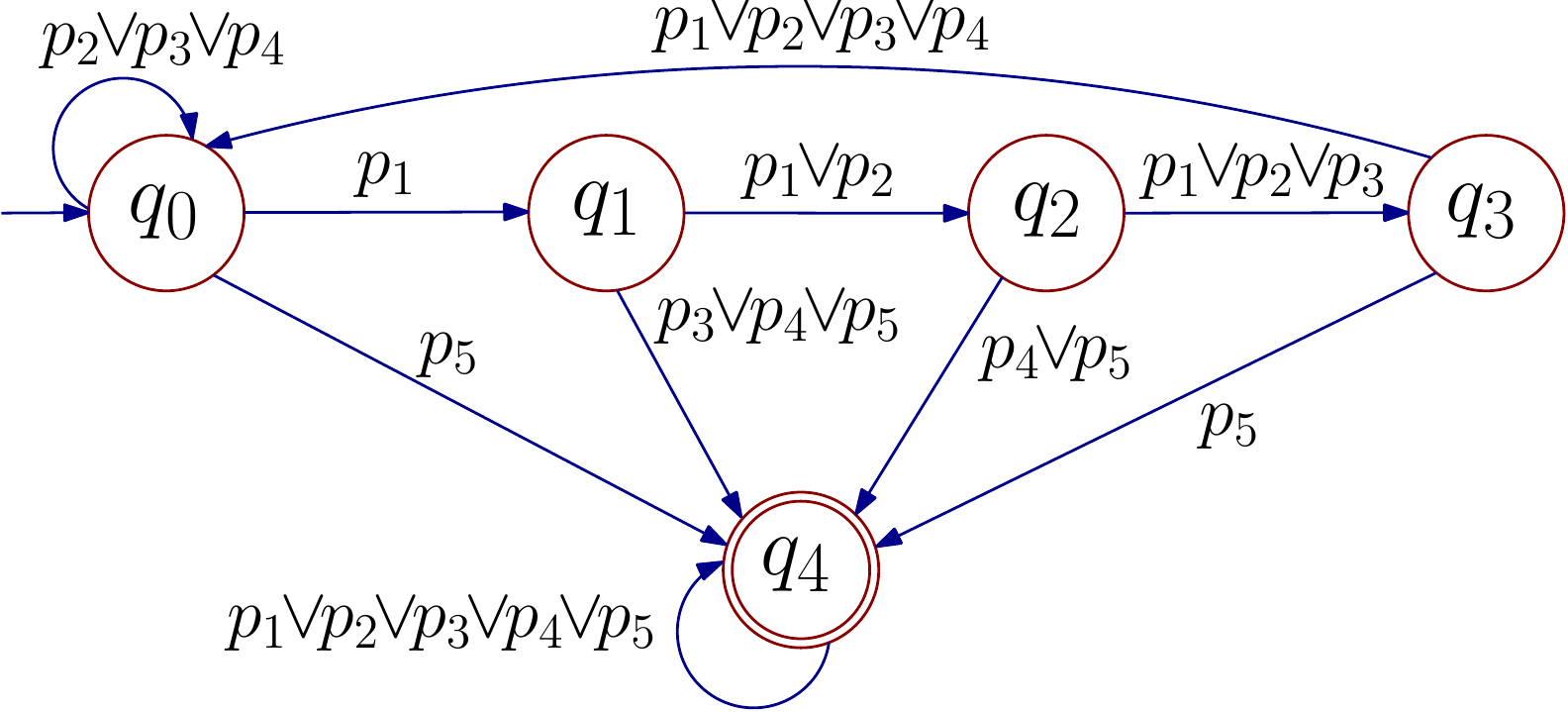}
	}
	\caption{\textbf{Left:} DFA $\mathcal{A}_E$, with accepting state $q_1$, alphabet $\Pi=\{p_1,p_2\}$, and labelling function $L:Y \rightarrow \Pi$ with $L(y)=p_1$ when $y \in [-0.5, 0.5]$, and $L(y)=p_2$ when $y \in (-\infty, -0.5)\cup (0.5,+\infty)$. ~\textbf{Right:} DFA $\mathcal{A}_N$, with accepting state $q_4$, alphabet $\Pi=\{p_1,p_2,p_3,p_4,p_5\}$, and labelling function $L:Y \rightarrow \Pi$ with 
		$L(y)=p_1$ when $y \in [-0.3, 0.3]$,
		$L(y)=p_2$ when $y \in [-0.4,-0.3)\cup (0.3, 0.4]$, 
		$L(y)=p_3$ when $y \in [-0.45,-0.4)\cup (0.4, 0.45]$, 
		$L(y)=p_4$ when $y \in [-0.5,-0.45)\cup (0.45, 0.5]$, and
		$L(y)=p_5$ when $y \in (-\infty,-0.5)\cup (0.5, +\infty)$.}\label{fig:DFA}
\end{figure}

{\bf Modeling and safety properties:}
By employing the feedback linearization technique proposed in~\cite{Ghaffari2021Analytical}, the relative motion between the quadrotor and the ground vehicle on $N$ and $E$ axes (see Figure~\ref{fig:experiment_setup}) can be modeled as:
\begin{align}
\!\!\!\!\!\left\{\hspace{-0.15cm}\begin{array}{l}
x_{\textsf i}(k+1) \!=\! Ax_{\textsf i}(k)+Bu_{\textsf i}(k)+Dw_{\textsf i}(k)+R\varsigma_{\textsf i}(k),\\
y_{\textsf i}(k)=Cx_{\textsf i}(k),\quad \quad \quad \quad \quad \quad k\in\mathbb N, \textsf i \in \{N,E \}, \end{array}\right.\label{quadrotor}
\end{align}
where 
$A \!=\! \begin{bmatrix}\begin{smallmatrix}1\,&\Delta t\\0\,&1\end{smallmatrix}\end{bmatrix}$,
$B \!=\! [\frac{\Delta t^2}{2};\Delta t]$, $D = -B$, and $C = [1;0]^{\top}$,
with $\Delta t = 0.1s$ being the sampling time, and $R \!=\! \begin{bmatrix}\begin{smallmatrix}0.004\,&0\\0\,&0.045\end{smallmatrix}\end{bmatrix}$ being obtained through experimental trials on our physical test-bed.
Here, for $\textsf i \in \{N,E\}$, $x_{\textsf i}(k) := [x_{\textsf i r}(k);$ $v_{\textsf i r}(k)]$ with $x_{\textsf i r}(k)$ and $v_{\textsf i r}(k)$ being the relative position and relative velocity between the quadrotor and the vehicle on $\textsf i$ axis, respectively;
$u_{\textsf i}(k)\in[-2.5,2.5]$ (m/s$^2$) denotes the acceleration of the quadrotor on $\textsf i$ axis as the control input;
$w_{\textsf i}(k)\in[-0.6,0.6]$ (m/s$^2$) denotes the acceleration of the vehicle on $\textsf i$ axis as the adversary input; 
$\varsigma_{\textsf i}(k)$ is a standard Gaussian random variable;
and $y_{\textsf i}(k)$ is the output.
Within 1 min (time horizon $H=600$), the following safety properties are desired:
(1) $(\mathcal{A}_E,H)$: $y_{E}$ should be within $[-0.5,0.5]$ (m);
(2) $(\mathcal{A}_N,H)$: $y_{N}$ should be within $[-0.5,0.5]$ (m); additionally, if $y_{N}$ reaches $[-0.3,0.3]$ (m) at any time instant $k$, then $y_{N}$ should be within $[-0.4,0.4]$ (m) at time instant $k+1$ and within $[-0.45,0.45]$ (m) at time instant $k+2$, instead of $[-0.5,0.5]$ (m).
The DFAs $\mathcal{A}_E$ and $\mathcal{A}_N$ are depicted in Figure~\ref{fig:DFA}.
Next, we proceed with designing the Safe-visor architecture with respect to $(\mathcal{A}_E,H)$ and $(\mathcal{A}_N,H)$, denoted by $sva_{E}$ and $sva_{N}$, respectively.

{\bf Construction of Safe-visor architecture:}
To construct the safety advisor as introduced in Section~\ref{sec:supervisor}, we apply the results in~\cite[Section 4]{Zhong2021Automata} for building the finite abstraction of the model as in~\eqref{quadrotor}.
Concretely, we select $X = [-0.5,0.5]\times[-0.4,0.4]$ and partition it uniformly with grid cells whose sizes are $(0.02,0.02)$.
Then, we uniformly divide the input set $[-2.5,2.5]$ for the quadrotor and the input set $[-0.6,0.6]$ for the ground vehicle with 25 and 12 cells, respectively.
By employing the results in~\cite[Section 4.3]{Zhong2021Automata}, the finite abstraction is $(\epsilon,\delta)$-stochastically simulated by the original model 
with respect to the relation 
$\mathscr{R} := \big\{(x,\hat{x})\,|\,(x-\hat{x})^{\top}M(x-\hat{x})\leq \epsilon^2\big\}$, and 
$\mathscr{R}_w :=\big\{(w,\hat{w})\,|\,(w$ $-\hat{w})^{\top}(w-\hat{w})\leq\tilde{\epsilon}^2\big\}$,
with $\delta = 0$, $\epsilon = 0.0674$, $\tilde{\epsilon}= 0.05$, and $M = \begin{bmatrix}\begin{smallmatrix}1.4632\ &0.1757\\0.1757\ &0.0666\end{smallmatrix}\end{bmatrix}$,
when the interface function  
$u:=K(x$ $-\hat{x})+\hat{u}$
is applied with $K\!=\!\begin{small}\begin{bmatrix}-16.66;-4.83\end{bmatrix}\end{small}^{\top}$ (cf. Figure~\ref{fig1:sa}), and $\hat{U}:=\{\hat{u}\in U|0.12\leq\hat{u}\leq 0.12\}$ is used for building the safety advisor.
With the finite abstraction and the ($\epsilon,\!\delta$)-approximate probabilistic relation in hand, we are ready to synthesize the safety advisors for $sva_{E}$ and $sva_{N}$.

After the safety advisors are constructed offline, one can readily implement the supervisors leveraging the results in Theorem~\ref{theorem:guarantee for wcv}, for which checking the non-emptiness of the set $U_f$ in~\eqref{eq:Uf} at run-time is necessary (cf. Proposition~\ref{PR_for_UC}).
Consider the current state $x(k)$ of the original system, $\hat{x}(k)$ in the current state of the Safe-visor architecture, and input $u_{uc}(k)$ provided by the unverified controller.
The set $U_f$ is not empty if there exists $\hat{u}\in\hat{U}$ such that 
\begin{align}
\lVert Ax(k)&+Bu_{uc}(k)+Dw(k)+R\varsigma(k)-(A\hat{x}(k)+B\hat{u}+D\hat{w}(k)+R\hat{\varsigma}(k)) \rVert_M\leq \epsilon,\label{eq:check1}
\end{align}
holds for all $\varsigma\in \mathbb{R}^2$, with $\lVert\bar{x}\rVert_M:=\sqrt{\bar{x}^TM\bar{x}}$.
By setting $\hat{\varsigma}(k) = \varsigma(k)$,~\eqref{eq:check1} holds if one has
$\lVert \varphi - B\hat{u}\rVert_M\leq \epsilon - \gamma$,
with $\varphi := A(x(k)-\hat{x}(k))+Bu_{uc}(k)$ and $\gamma := \max_{\beta\in\Delta}\lVert\beta\rVert_M + \max_{(w,\hat{w})\in\mathscr{R}_w }\lVert D(w-\hat{w})\rVert_M=0.0152$, where $\Delta$ denotes the set of all possible quantization errors introduced by discretization of the original state set~\cite[eq. (4.11)]{Zhong2021Automata}.
Since $\varphi$ can readily be computed at run-time, one can find out whether there exists $\hat{u}\in\hat{U}$ such that~\eqref{eq:check1} holds efficiently.

\begin{small}
	\begin{table*}[]
		\centering
		\caption{Results of simulation and experiment on the physical test-bed.}\label{tab1}
		\begin{tabular}{|l|l|l|l|l|l|l|}
			\hline
			Safety properties &
			\begin{tabular}[c]{@{}l@{}}Satisfaction \\ rate (with \\$sva_E$/$sva_N$)\end{tabular} &
			\begin{tabular}[c]{@{}l@{}}Acceptance \\ rate \end{tabular} &
			\begin{tabular}[c]{@{}l@{}}Satisfaction \\ rate (without \\$sva_E$/$sva_N$)\end{tabular} &
			\begin{tabular}[c]{@{}l@{}}Average execution\\ time (ms)\end{tabular} &
			\begin{tabular}[c]{@{}l@{}}Standard deviation of \\ the execution time (ms)\end{tabular} \\ \hline
			$(\mathcal{A}_E,H)$ (Simulation) & 100\% & 70.51\% & 0\%  & 2.3719 & 1.4570 \\ \hline
			$(\mathcal{A}_N,H)$ (Simulation) & 100\% & 70.03\% & 0\% & 2.3826 & 1.4503 \\ \hline
			$(\mathcal{A}_E,H)$ (Test-bed) & N.A. & 2.50\% & N.A.  & 4.2301 & 2.1056 \\ \hline
			$(\mathcal{A}_N,H)$ (Test-bed) & N.A. & 8.00\% & N.A.  & 3.3957 & 2.7630 \\ \hline
		\end{tabular}
	\end{table*}
\end{small}
{\bf Simulation:}
As for the simulation, we initialize the system at $x_E = x_N=[0.2;0.2]$ and set the maximal tolerable probability of violation for $sva_E$ and $sva_N$ as $\eta_E=\eta_N =0.01$.
In the simulation, the ground vehicle randomly selects inputs on both direction following a uniform distribution\footnote{Here, the ground vehicle does not select adversarial inputs rationally since it is challenging to obtain closed-form solutions for such adversarial strategies.
	Meanwhile, the probabilistic guarantees provided are valid regardless of how the ground vehicle chooses inputs (cf. Remark~\ref{rem:asy_info}).}$\!$ .
For demonstration purposes, we utilize an unverified controller for the quadrotor that selects inputs from its feasible set following a uniform distribution.
The simulation results are summarized in Table~\ref{tab1}, with~\emph{acceptance rate} being the average percentage of inputs from the unverified controller being accepted among different runs in the simulation.
One can verify that the desired lower bound of safety probability specified by $\eta_E$ and $\eta_N$ are respected. 
Simultaneously, the unverified controller can still be used whenever it does not endanger the safety of the system.

\begin{figure}
	\centering
	\includegraphics[width=8cm]{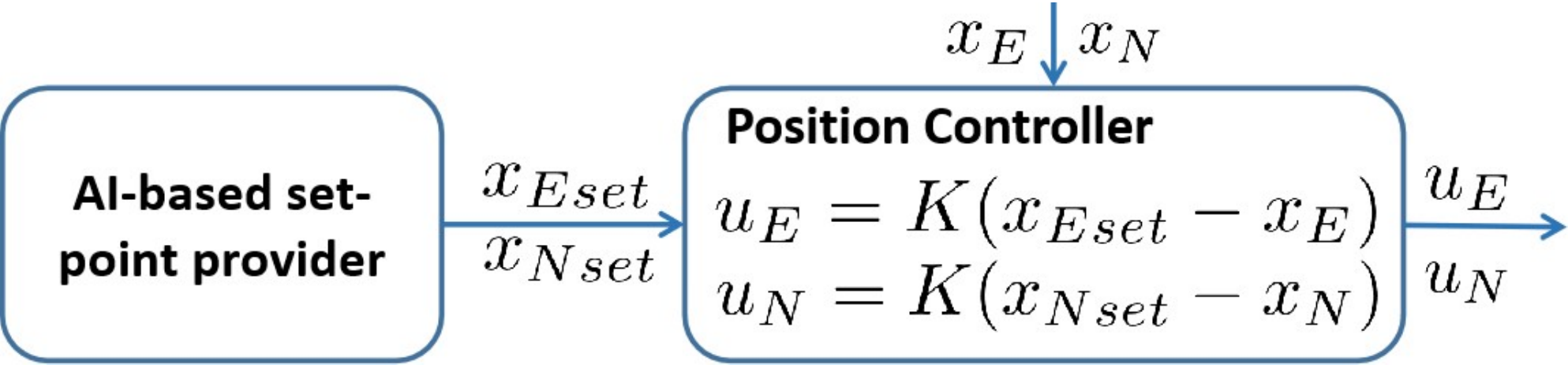}
	\caption{AI-based unverified controller employed in real-world experiments.} \label{fig1:AIcontroller}
\end{figure}

\begin{figure}
	\centering
	\includegraphics[width=9cm]{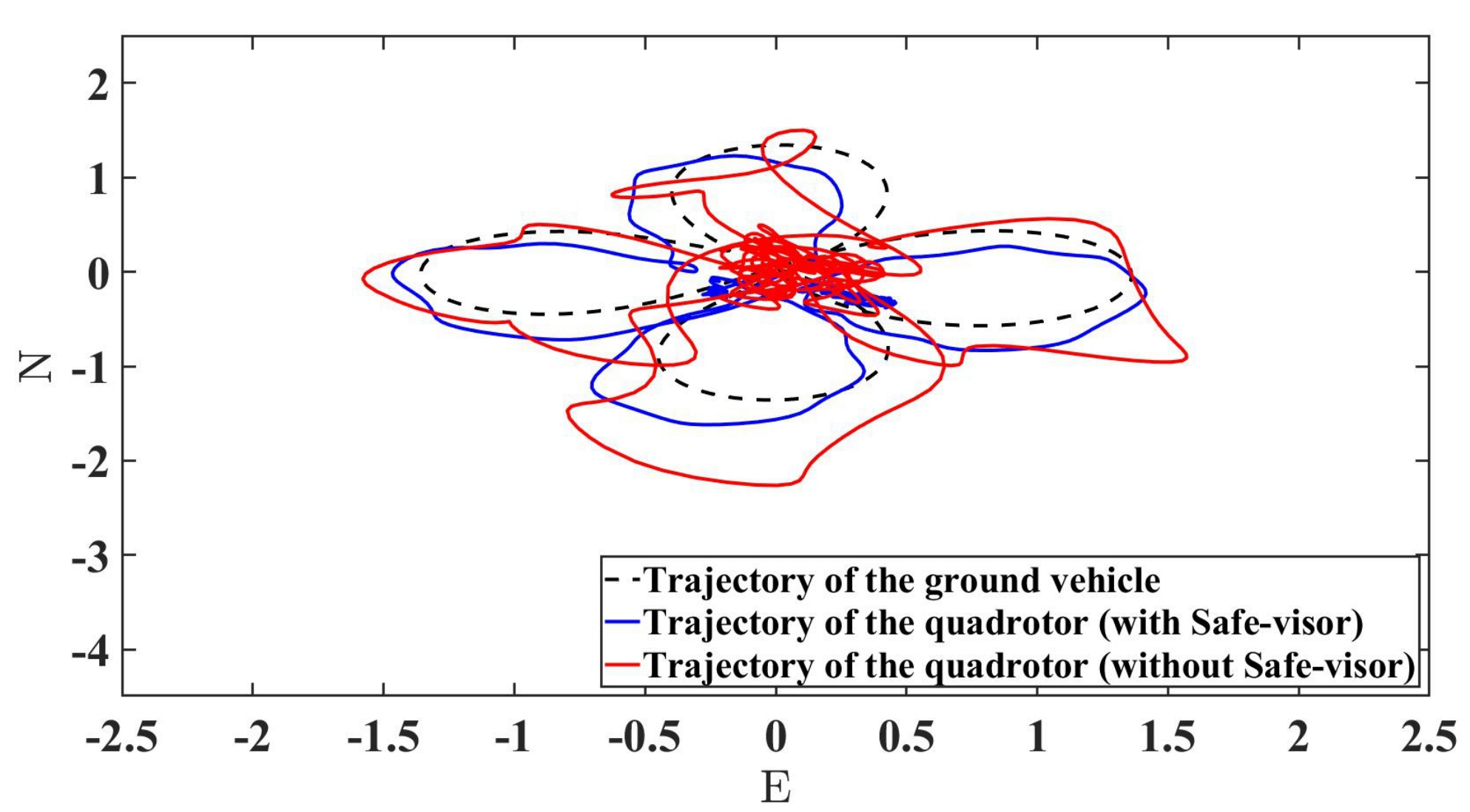}
	\caption{Trajectories of the quadrotor helicopter and the ground vehicle.}\label{fig:traj} 
\end{figure}

\begin{figure}
	\centering
	\includegraphics[width=9cm]{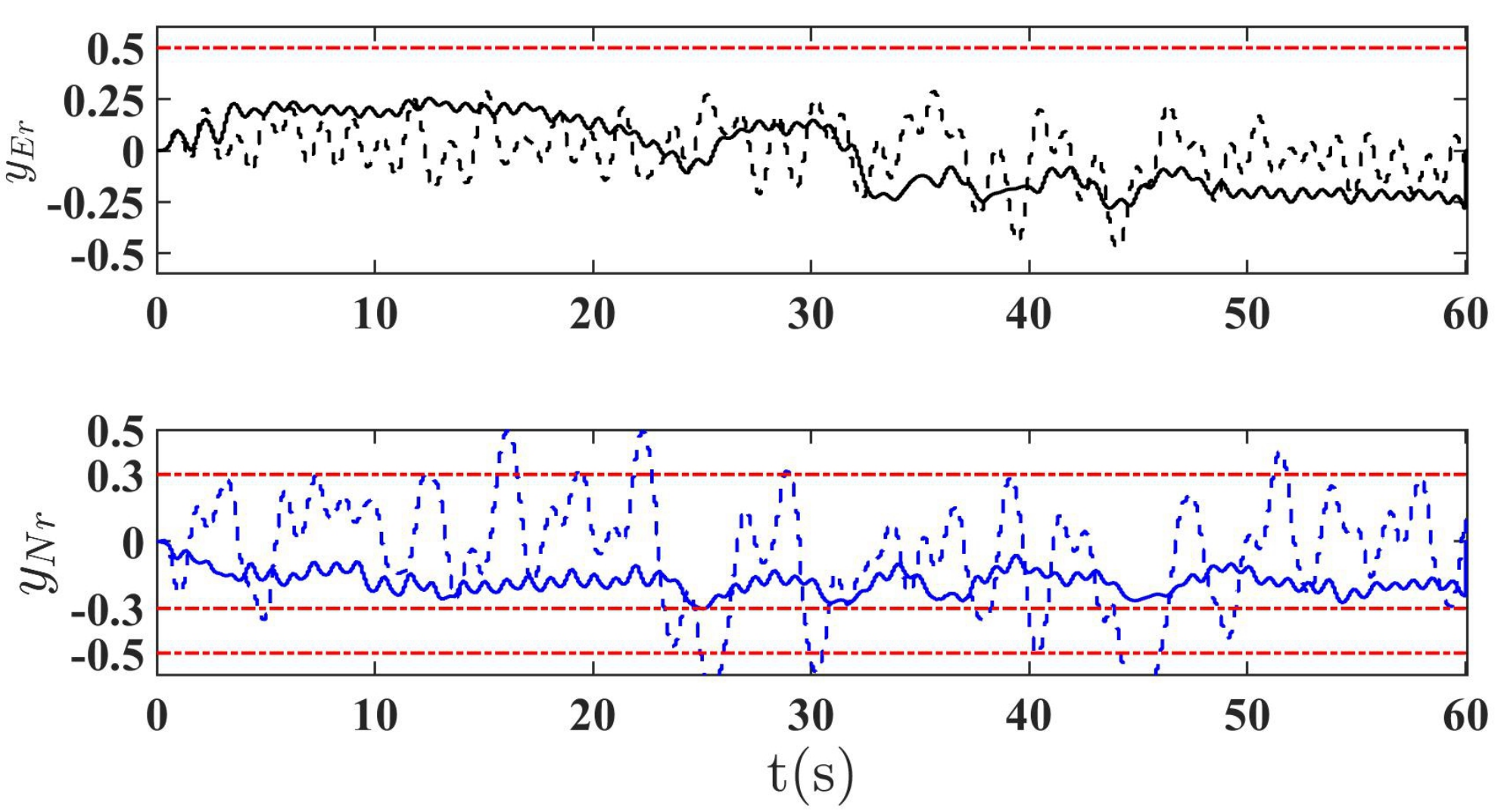}
	\caption{Evolution of $y_E$ and $y_N$ with (solid lines) and without (dashed lines) using $sva_E$ and $sva_N$.}\label{fig:eve} 
\end{figure}

{\bf Experiments on the physical test-bed:}
For demonstration purposes, we use an AI-based controller, as depicted in Figure~\ref{fig1:AIcontroller}, to control the quadrotor, with $K=\begin{small}\begin{bmatrix}1.4781;1.7309\end{bmatrix}\end{small}^{\top}\!$.
We initialized the system with $x_E = x_N=[0;0]$ and let the ground vehicle track a clover trajectory within our test field.
Here, the AI-based set-point provider is trained using deep reinforcement learning with deep deterministic policy gradient (DDPG) algorithms~\cite{Lillicrap2016Continuous}.
We are omitting details for the training procedure since designing and improving the performance of (AI-based) unverified controllers are out of the scope of this paper.
The results are summarized in Table~\ref{tab1} and depicted in Figure~\ref{fig:traj} and~\ref{fig:eve}.
Note that the AI-based controller behaves worse on the physical test-bed than in simulation due to the  mismatch between the model that is used for training and the physical system.
One can verify that the desired safety property is violated when only applying the unverified controller (see the bottom part of Fig~\ref{fig:eve}, in which $y_{Nr}$ left the region $[-0.5,0.5]$).
Meanwhile, by leveraging $sva_E$ and $sva_N$, the desired safey properties are enforced,
while the unverified controller can still be employed.


\section{Conclusions}\label{sec:discussion}
In this paper, we proposed, for the first time, the construction of a Safe-visor architecture for sandboxing unverified controllers in two-players non-cooperative stochastic games with continuous state and input sets. 
This architecture consists of 1) a supervisor that checks the inputs provided by the unverified controller and rejects them whenever they endanger the overall safety of the systems; 2) a safety advisor that provides inputs maximizing the probability of satisfying the desired safety properties if the unverified controller is rejected.
Both components are constructed using abstraction-based approaches, and formal safety guarantees are provided based on the ($\epsilon,\delta$)-approximate probabilistic relation between the original game and its finite abstraction. 
The compromise between safety and functionality is achieved by setting a maximal tolerable probability of violating the desired safety properties.
Finally, the effectiveness of our results is demonstrated through simulations and experiments on a physical test-bed for a quadrotor helicopter.

\section{Acknowledgment}
This work was supported in part by the H2020 ERC Starting Grant AutoCPS (grant agreement No 804639) and by an Alexander von Humboldt Professorship endowed by the German Federal Ministry of Education and Research.

\bibliographystyle{my-elsarticle-num}
%\biboptions{sort&compress}
%\bibliography{autosam}           % and a bib file to produce the     
\bibliography{mybib}  
% and a bib file to produce the 
% bibliography (preferred). The
% correct style is generated by
% Elsevier at the time of printing.

\appendix
\section{Proof for Theorem~\ref{theorem:guarantee for wcv}}\label{proof}
Before showing the proof for Theorem~\ref{theorem:guarantee for wcv}, some additional definitions and lemmas are required.
First, we define the \emph{n-steps reachable state set} of a DFA as follows.
\begin{definition}
	\emph{(n-steps Reachable State Set)}
	Consider a gDTSG $\mathfrak{D}= (X,U,W,X_0,T,Y,h)$ and a DFA $\mathcal{A} = (Q, q_0, \Pi,\tau, F)$ with a labelling function $L:Y\rightarrow\Pi$. 
	An \emph{n-steps reachable state set} $\tilde{Q}_n(x_0)$ of $\mathcal{A}$ with respect to an initial state $x_0\in X_0$ is recursively defined as
	\begin{align*}
	\tilde{Q}_0(x_0) &= \Big\{q\in Q\,\big|\,q=\tau(q_0,L\circ h(x_0))\Big\},\\
	\tilde{Q}_n(x_0)&= \Big\{q\in Q\,\big|\,\exists q'\in \tilde{Q}_{n-1}(x_0), \sigma \in \Pi\ s.t.\ q=\tau(q',\sigma)\Big\},
	\end{align*}
	with $n\in \N_{>0}$.
\end{definition} 
Additionally, we need the definitions of Markov policy, Product gDTSG, and a cost-to-go function, which are borrowed from~\cite{Zhong2021Automata}.

\begin{definition}\label{def_mp}
	\emph{(Markov Policy)} Consider a gDTSG $\mathfrak{D} =(X,U,$ $W,X_0,T,Y,h)$. A \textit{Markov policy} $\rho$ defined over the time horizon $[0,H-1]\subset\N$ for Player~\uppercase\expandafter{\romannumeral1} is a sequence $\rho\!=\!(\rho_{0},\rho_{1},\ldots,\rho_{H-1})$ of universally measurable maps $\rho_{k}:X\rightarrow\mathbf{P}(U,\mathcal{B}(U))$, with
	\begin{equation*}
	\rho_{k}(U\big | x(k)) =1.
	\end{equation*} 
	Similarly, 	
	a \textit{Markov policy} $\lambda$ for Player~\uppercase\expandafter{\romannumeral2} is a sequence $\lambda\,=\,(\lambda_{0},\,\lambda_{1},\,\ldots,\lambda_{H-1})$ of universally measurable maps $\lambda_{k}:\,X\times U\,\rightarrow\,\mathbf{P}(W,\mathcal{B}(W))$, with
	\begin{equation*}
	\lambda_{k}(W\big | x(k),u(k)) =1,
	\end{equation*}
	for all $k\in[0,H-1]$.
	We use $\mathcal{P}$ and $\Lambda$ to denote the set of all Markov policies for Players~\uppercase\expandafter{\romannumeral1} and~\uppercase\expandafter{\romannumeral2}, respectively. 
	Moreover, we denote by $\mathcal{P}^H$ and $\Lambda^H$ the set of all Markov policies for Players~\uppercase\expandafter{\romannumeral1} and~\uppercase\expandafter{\romannumeral2} within time horizon $[0,H-1]$, respectively.
\end{definition}

\begin{definition}\label{def:product_gmdp_dfa} \emph{(Product gDTSG)}
	Consider a gDTSG $\mathfrak{D} =(X,U,W,X_0,T,Y,h)$, a DFA $\mathcal{A} = (Q, q_0, \Pi, \tau, F)$, and a labelling function $L: Y\rightarrow \Pi$ as in Definition~\ref{def:sactisfaction_DFA}.
	The product of $\mathfrak{D}$ and $\mathcal{A}$ is a gDTSG defined as
	\begin{equation*}
	\mathfrak{D}\otimes\mathcal{A} = \{\bar{X},\bar{U},\bar{W},\bar{X}_0,\bar{T},\bar{Y},\bar{h}\},
	\end{equation*}
	where
	$\bar{X}:=X\times Q$ is the state set;
	$\bar{U}:= U$ is the input set for Player~\uppercase\expandafter{\romannumeral1}; 
	$\bar{W}:= W$ is the input set for Player~\uppercase\expandafter{\romannumeral2};	
	$\bar{X}_0$ is the initial state set, with $\bar{x}_0:=(x_0,\bar{q}_0)\in\bar{X_0}$, $x_0\in X_0$ and
	\begin{equation}
	\bar{q}_0 = \tau(q_0,L\circ h(x_0));\label{eq:compute_initial_q0}
	\end{equation}
	$\bar{T}(\mathsf dx'\times\{q'\}|x,q,u,w)$ is the stochastic kernel that assigns for any $(x,q)\in\bar{X}$, $u\in\bar{U}$, and $w\in\bar{W}$ the probability $\bar{T}(\mathsf dx'\times\{q'\}|x,q,u,w)=T(\mathsf dx'|x,u,w)$ when $q' =\tau(q,L\circ h(x'))$, and $\bar{T}(\mathsf dx'\times\{q'\}|x,q,w,u)=0$, otherwise; $\bar{Y}:=Y$ is the output set and $\bar{h}(x,q):=h(x)$ is the output map.
\end{definition}

\begin{definition}\label{def:value_f_wcv}
	Given a Markov policy $\rho=(\rho_{0},\rho_{1},\ldots,\rho_{H-1})$ for Player~\uppercase\expandafter{\romannumeral1} and $\lambda=(\lambda_{0},\lambda_{1},\ldots,\lambda_{H-1})$ for Player~\uppercase\expandafter{\romannumeral2} of $\widehat{\mathfrak{D}}\otimes\mathcal{A}$, we define a cost-to go function $\ul{V}^{\rho,\lambda}_n:\hat{X}\times Q\rightarrow [0,1]$, which is initialized by $\ul{V}_0^{\rho,\lambda}(\hat{x},q)=1$ when $q\in F$, and $\ul{V}_0^{\rho,\lambda}(\hat{x},q)=0$, otherwise, and recursively computed as
	\begin{align}
	\ul{V}_{n+1}^{\rho,\lambda}(\hat{x},q):=
	&\left\{
	\begin{aligned} 
	&(1-\delta)\sum_{\hat{x}'\in \hat{X}}\ul{V}^{\rho,\lambda}_{n}(\hat{x}',\bar{q}(\hat{x}',q))\hat{T}(\hat{x}'|\hat{x},\hat{u},\hat{w})+\delta, \text{ if } q\notin F;\\
	& \quad \quad \quad \quad 1, \quad \quad \quad \quad \quad\, \quad \quad \quad\quad \quad \quad\quad \quad \quad \text{ if } q\in F,
	\end{aligned}\right.\label{eq:value_gen_vio_ab}
	\end{align}
	where $\hat{u}=\rho_{H-n-1}(\hat{x},q)$, $\hat{w}=\lambda_{H-n-1}(\hat{x},q,\hat{u})$, and
	\begin{equation}\label{eq:overline_pg}
	\overline{q}(\hat{x}',q) = \mathop{\arg\max}_{q'\in Q'_{\epsilon}(\hat{x}')}\ul{V}^{\rho,\lambda}_{n}(\hat{x}',q'),
	\end{equation}
	with $Q'_{\epsilon}(\hat{x}')$ as in Definition~\ref{Vstar}.
\end{definition}
Additionally, the worst-case adversarial policy $\lambda^*(\rho)$ for Player~\uppercase\expandafter{\romannumeral2} with respect to the Markov policy $\rho$ for Player~\uppercase\expandafter{\romannumeral1} can be computed as
\begin{align}
\lambda^*_{H-n-1}(\rho)\in\max_{\lambda_{H-n-1}\in \Lambda}\big((1-\delta)\sum_{\hat{x}'\in \hat{X}}\ul{V}^{\rho,\lambda^*(\rho)}_{n}(\hat{x}',\overline{q}(\hat{x}',q))\hat{T}(\hat{x}'|\hat{x},\hat{u},\hat{w})+\delta\big)\label{eq:policy_min_worst_case},
\end{align}
for all $n\in[0,H-1]$, with $\hat{u}=\rho_{H-n-1}(\hat{x},q)$, and $\hat{w}=\lambda_{H-n-1}(\hat{x},q,\hat{u})$.

With Definition~\ref{def:value_f_wcv} in hand, we have the following lemma, which is required for showing the results of Theorem~\ref{theorem:guarantee for wcv}.
\begin{lemma}\label{lem:cosafeLTL1minus}
	Given a gDTSG $\mathfrak{D}=(X,U,W,X_0,T,Y,h)$ and its finite abstraction $\widehat{\mathfrak{D}}= (\hat{X},\hat{U}, \hat{W},\hat{X}_0, \hat{T}, Y,\hat{h})$ with $\widehat{\mathfrak{D}}\preceq^{\delta}_{\epsilon}\mathfrak{D}$, and a desired safety property $(\mathcal{A},H)$ with $\mathcal{A}=(Q, q_0, \Pi,$ $\tau, F)$.
	Given a Markov policy $\rho\,=\,(\rho_0, \rho_1,\ldots,$ $\rho_{H-1})$ over the product gDTSG $\widehat{\mathfrak{D}}\otimes \mathcal{A}$ within the time horizon $[0,H]$, one has
	\begin{equation}\label{eq:help2}
	1-\ul{V}_{n+1}^{\rho,\lambda^*(\rho)}(\hat{x},q)= (1-\delta)\min_{\lambda_{H-n-1}\in \Lambda}\sum_{\tilde{x}\in\hat{X}'_{-\epsilon}(q)}\Big(1-\ul{V}_{n}^{\rho,\lambda^*(\rho)}(\tilde{x},\bar{q}(\tilde{x},q))\Big)\hat{T}(\tilde{x}\,\big|\,\hat{x},\hat{u},\hat{w}),\\
	\end{equation}
	with $q\notin F$, $\hat{u}=\rho_{H-n-1}(\hat{x},q)$, $\hat{w}=\lambda_{H-n-1}(\hat{x},q,\hat{u})$, $\ul{V}_{n+1}^{\rho,\lambda^*(\rho)}(\hat{x},q)$ as in~\eqref{eq:value_gen_vio_ab}, $\lambda^*(\rho)$ as in~\eqref{eq:policy_min_worst_case}, $\bar{q}$ as in~\eqref{eq:overline_pg}, and $\hat{X}'_{-\epsilon}(q)$ as in Definition~\ref{def:History-based Supervisor_wcv}.
\end{lemma}	
The proof of Lemmas~\ref{lem:cosafeLTL1minus} can readily be derived with the help of~\eqref{eq:value_gen_vio_ab} and \eqref{eq:overline_pg}. 
Now, we are ready to show the results for Theorem~\ref{theorem:guarantee for wcv}.

{\bf Proof of Theorem~\ref{theorem:guarantee for wcv}}
For the sake of clarity of the proof, we define
\begin{align*}
&r(\hat{x}(k),q(k),\hat{w}(k))=1-\max_{\lambda_{k}\in \Lambda}\sum_{\hat{x}(k+1)\in \hat{X}}\!\!\!\ul{V}_{*,H-k-1}\Big(\hat{x}(k+1),\bar{q}_*\big(\hat{x}(k+1),q(k)\big)\Big)\hat{T}\big(\hat{x}(k+1)\,\big|\,\hat{x}(k),\hat{u}(k),\hat{w}(k)\big),
\end{align*}
for $k\in[0,H-1]$ with $\ul{V}_{*,H-k-1}$ and $\underline{q}^*$ as in Definition~\ref{Vstar}, $\hat{u}(k)=\rho'_k(\hat{x}(k),q(k))$, and $\hat{w}(k)=\lambda_k(\hat{x}(k),q(k),\hat{u}(k))$ and
\begin{align*}
g(\hat{x}(z-1),q(z-1),\hat{w}(z-1))=\hat{T}\big(\hat{x}(z)\,\big|\,\hat{x}(z-1),\rho'_{z-1}(\hat{x}(z-1),q(z-1)),\hat{w}(z-1)\big),
\end{align*}	
for $z\in[0,k]$.
Consider an initial state $(\hat{x}_0,\bar{q}_0)$, with $\hat{x}_0\in\hat{X}_0$.
By leveraging~\eqref{eq:help2}, we expand out $\ul{V}^{\rho',\lambda^*(\rho')}_H(\hat{x}_0,\bar{q}_0)$ up to the time instant $k\in[0,H-1]$ as 
\begin{small}
	\begin{align}
	&1-\ul{V}^{\rho',\lambda^*(\rho')}_H(\hat{x}_0,\bar{q}_0)\nonumber\\
	=&(1-\delta)^{k+1}\min_{\lambda_{0}\in \Lambda}\sum_{\hat{x}(1)\in \hat{X}'_{-\epsilon}(q(0))}\!\!\!\!\Big(\min_{\lambda_{1}\in \Lambda} \sum_{\hat{x}(2)\in \hat{X}'_{-\epsilon}(q(1))}\!\!\!\!\!\!\!\!\Big(\ldots\Big(\min_{\lambda_{k-2}\in \Lambda}\sum_{\hat{x}(k-1)\in \hat{X}'_{-\epsilon}(q(k-2))}\Big(\min_{\lambda_{k-1}\in \Lambda}\sum_{\hat{x}(k)\in \hat{X}'_{-\epsilon}(q(k-1))}\!\!\!\!\!\!\!\!\!\!r\big(\hat{x}(k),q(k),\hat{w}(k)\big)\nonumber\\
	&\times g\big(\hat{x}(k-1),q(k-1),\hat{w}(k-1)\big)\Big)g\big(\hat{x}(k-2),q(k-2),\hat{w}(k-2)\big)\Big)\!\ldots\!\Big)g\big(\hat{x}(1),q(1),\hat{w}(1)\big)\Big)g\big(\hat{x}(0),q(0),\hat{w}(0)\big),\label{eq:proof2step1}
	\end{align}
\end{small}\noindent
with $\hat{w}(m):=\lambda_m(\hat{x}(m),\rho'_{m}(\hat{x}(m),q(m)))$, $m\in[0,k-1]$.
Firstly, by selecting 
\begin{small}
	\begin{align*}
	\hat{x}_*(k) := \mathop{\arg\min}_{\hat{x}(k)\in \hat{X}'_{-\epsilon}(q(k-1))}r(\hat{x}(k),q(k),\hat{w}(k)),
	\end{align*}
\end{small}\noindent
with $q(k-1)\in\tilde{Q}_{k-1}(x_0)$, $\hat{X}'_{-\epsilon}(q(k-1))$ as in Definition~\ref{def:History-based Supervisor_wcv}, and $q_*(k)=\bar{q}(q(k-1),\hat{x}_*(k))$ with $\bar{q}$ as in \eqref{eq:overline_pg}, one has
\begin{small}
	\begin{align*}
	&\min_{\lambda_{k-1}\in \Lambda}\sum_{\hat{x}(k)\in \hat{X}'_{-\epsilon}(q(k-1))}\!\!\!\!\!\!\!\!\!\!\!\!\!\!\!r(\hat{x}(k),q(k),\hat{w}(k))g(\hat{x}(k-1),q(k-1),\hat{w}(k-1))\\
	\geq &\ r(\hat{x}_*(k),q_*(k),\hat{w}(k))\min_{\lambda_{k-1}\in \Lambda}\sum_{\hat{x}(k)\in \hat{X}'_{-\epsilon}(q(k-1))}\!\!\!\!\!\!\!\!\!\!\!\!\!\!\!g(\hat{x}(k-1),q(k-1),\hat{w}(k-1)),
	\end{align*}
\end{small}\noindent
with $\hat{w}(k-1)=\lambda_{k-1}\Big(\hat{x}(k-1),\rho'_{k-1}(\hat{x}(k-1),q(k-1))\Big)$.
Hence, proceed with~\eqref{eq:proof2step1}, one gets 
\begin{small}
	\begin{align}
	&1-\ul{V}^{\rho',\lambda^*(\rho')}_H(\hat{x}_0,\bar{q}_0)\nonumber\\
	\geq&(1-\delta)^{k+1}\min_{\lambda_{0}\in \Lambda} \sum_{\hat{x}(1)\in \hat{X}'_{-\epsilon}(q(0))}\!\!\!\!\!\!\!\!\Big( \min_{\lambda_{1}\in \Lambda}\sum_{\hat{x}(2)\in \hat{X}'_{-\epsilon}(q(1))}\!\!\!\!\Big(\ldots\Big(\min_{\lambda_{k-2}\in \Lambda}\!\!\sum_{\hat{x}(k-1)\in \hat{X}'_{-\epsilon}(q(k-2))}\Big(\min_{\lambda_{k-1}\in \Lambda}\!\!\!\sum_{\hat{x}(k)\in \hat{X}'_{-\epsilon}(q(k-1))}\!\!\!\!\!\!\!\!\!\!\!\!\!\!\!\!g\big(\hat{x}(k-1),q(k-1),\nonumber\\
	&\hat{w}(k-1)\big)\Big) g\big(\hat{x}(k-2),q(k-2),\hat{w}(k-2)\big)\Big)\ldots\Big)g\big(\hat{x}(1),q(1),\hat{w}(1)\big)\Big)g\big(\hat{x}(0),q(0),\hat{w}(0)\big)r\big(\hat{x}_*(k),q_*(k),\hat{w}(k)\big).\label{eq:proof2step2}
	\end{align}
\end{small}\noindent
with $\hat{w}(m):=\lambda_m(\hat{x}(m),\rho'_{m}(\hat{x}(m),q(m)))$, $m\in[0,k-1]$.
Secondly, we select 
\begin{equation}\label{proof2opt1}
\hat{x}_*(k-1) = \mathop{\arg\min}_{\hat{x}(k-1)\in \hat{X}'_{-\epsilon}(q(k-2))}\min_{\lambda_{k-1}\in \Lambda}\sum_{\hat{x}(k)\in \hat{X}'_{-\epsilon}(q(k-1))}g\big(\hat{x}(k-1), q(k-1),\hat{w}(k-1)\big),
\end{equation}
with $\hat{w}(k-1)=\lambda_{k-1}\Big(\hat{x}(k-1),\rho'_{k-1}(\hat{x}(k-1),q(k-1))\Big)$, $q(k-2)\in \tilde{Q}_{k-2}(x_0)$, and 
\begin{equation}\label{proof2opt2}
q_*(k-1)=\bar{q}(q(k-2),\hat{x}_*(k-1)),
\end{equation}
with $\bar{q}$ as in \eqref{eq:overline_pg}.
Then, one has
\begin{small}
	\begin{align*}
	&\min_{\lambda_{k-2}\in \Lambda}\sum_{\hat{x}(k-1)\in \hat{X}'_{-\epsilon}(q(k-2))}\Big(\min_{\lambda_{k-1}\in \Lambda}\sum_{\hat{x}(k)\in \hat{X}'_{-\epsilon}(q(k-1))}\!\!\!\!\!\!\!\!\!\!\!\!\!\!\!\!g\big(\hat{x}(k-1),q(k-1),\hat{w}(k-1)\big)\Big) g\big(\hat{x}(k-2),q(k-2),\hat{w}(k-2)\big)\nonumber\\
	\geq&\quad \Big(\min_{\lambda_{k-1}\in \Lambda}\sum_{\hat{x}(k)\in \hat{X}'_{-\epsilon}(q_*(k-1))}\!\!\!\!\!\!\!\!\!\!\!\!\!\!\!\!g\big(\hat{x}_*(k-1),q_*(k-1),\hat{w}(k-1)\big)\Big) \min_{\lambda_{k-2}\in \Lambda}\sum_{\hat{x}(k-1)\in \hat{X}'_{-\epsilon}(q(k-2))}g\big(\hat{x}(k-2),q(k-2),\hat{w}(k-2)\big).
	\end{align*}
\end{small}
Thus, proceed from~\eqref{eq:proof2step2}, we have
\begin{small}
	\begin{align}
	&1-\ul{V}^{\rho',\lambda^*(\rho')}_H(\hat{x}_0,\bar{q}_0)\nonumber\\
	\geq&(1-\delta)^{k+1}\min_{\lambda_{0}\in \Lambda}\sum_{\hat{x}(1)\in \hat{X}'_{-\epsilon}(q(0))}\!\!\!\!\Big(\min_{\lambda_{1}\in \Lambda} \sum_{\hat{x}(2)\in \hat{X}'_{-\epsilon}(q(1))}\!\!\!\!\Big(\ldots\Big(\min_{\lambda_{k-2}\in \Lambda}\sum_{\hat{x}(k-1)\in \hat{X}'_{-\epsilon}(q(k-2))}g\big(\hat{x}(k-2),q(k-2),\hat{w}(k-2)\big)\Big)\ldots\Big)\nonumber\\
	&\times\, g\big(\hat{x}(1),q(1),\hat{w}(1)\big)\Big)g\big(\hat{x}(0),q(0),\hat{w}(0)\big)\Big(\min_{\lambda_{k-1}\in \Lambda}\sum_{\hat{x}(k)\in \hat{X}'_{-\epsilon}(q_*(k-1))}\!\!\!\!\!\!\!\!\!\!\!\!\!\!\!\!g\big(\hat{x}_*(k-1),q_*(k-1),\hat{w}(k-1)\big)\Big)r\big(\hat{x}_*(k),q_*(k),\hat{w}(k)\big).\label{eq:proof2step3}
	\end{align}
\end{small}\noindent
with $\hat{w}(m):=\lambda_m(\hat{x}(m),\rho'_{m}(\hat{x}(m),q(m)))$, $m\in[0,k-1]$.
For all $z\in[2,k-1]$, by choosing $x_*(z-1)$ similar to~\eqref{proof2opt1} and $q_*(z-1)$ similar to~\eqref{proof2opt2}, one obtains
\begin{small}
	\begin{align}
	&\min_{\lambda_{z-2}\in \Lambda}\sum_{\hat{x}(z-1)\in \hat{X}'_{-\epsilon}(q(z-2))}\Big(\min_{\lambda_{z-1}\in \Lambda}\sum_{\hat{x}(z)\in \hat{X}'_{-\epsilon}(q(z-1))}\!\!\!\!\!\!\!\!\!\!\!\!\!\!\!\!g\big(\hat{x}(z-1),q(z-1),\hat{w}(z-1)\big)\Big) g\big(\hat{x}(z-2),q(z-2),\hat{w}(z-2)\big)\nonumber\\
	\geq&\,\Big(\min_{\lambda_{z-1}\in \Lambda}\!\!\sum_{\hat{x}(z)\in \hat{X}'_{-\epsilon}(q_*(z-1))}\!\!\!\!\!\!\!\!\!\!\!\!\!\!\!\!g\big(\hat{x}_*(z-1),q_*(z-1),\hat{w}(z-1)\big)\Big)\min_{\lambda_{z-2}\in \Lambda}\!\!\sum_{\hat{x}(z-1)\in \hat{X}'_{-\epsilon}(q(z-2))}\!\!\!\!g\big(\hat{x}(z-2),q(z-2),\hat{w}(z-2)\big).\label{eq:proof2final}
	\end{align}
\end{small}\noindent
with $\hat{w}(m):=\lambda_m\Big(\hat{x}(m),\rho'_{m}(\hat{x}(m),q(m))\Big)$, $m\in\{z-1,z-2\}$.
Then, with~\eqref{eq:proof2final} for all $z\in[2,k-1]$ and~\eqref{eq:proof2step3}, one gets
\begin{small}
	\begin{align*}
	&1-\ul{V}^{\rho',\lambda^*(\rho')}_H(\hat{x}_0,\bar{q}_0)\\
	\geq\  &(1-\delta)^{k+1}\Big(\min_{\lambda_{0}\in \Lambda}\!\!\!\sum_{\hat{x}(1)\in \hat{X}'_{-\epsilon}(q(0))}\!\!\!\!\!\!\!\!\!\!\!\!\!g(\hat{x}(0),q(0),\hat{w}(0))\Big)\prod_{z=2}^{k}\Big(\min_{\lambda_{z-1}\in \Lambda}\!\!\!\sum_{\hat{x}(z)\in \hat{X}'_{-\epsilon}(q_*(z-1))}\!\!\!\!\!\!\!\!\!\!\!\!\!\!\!\!\!g(\hat{x}_*(z\!-\!1),q_*(z\!-\!1),\hat{w}(z\!-\!1))\Big)r(\hat{x}_*(k),q_*(k),\hat{w}(k)).
	\end{align*}
\end{small}\noindent
with $\hat{w}(m):=\lambda_m(\hat{x}(m),\rho'_{m}(\hat{x}(m),q(m)))$, $m\in[0,k-1]$.
Note that $\bar{\omega}'_k := \big(\hat{x}(0), q(0),\rho_0(\hat{x}(0),q(0)),\hat{x}_*(1), q_*(1),$ $\rho_1(\hat{x}_*(1),q_*(1))\ldots\,\hat{x}_*(k),q_*(k)\big)$ is one of history paths of the memory state of the safe-visor architecture up to the time instant $k$, and the supervisor as in Definition~\ref{def:History-based Supervisor_wcv} ensures that for all such history paths $\bar{\omega}_k$, we have
\begin{equation*}
(1-\delta)^{k+1}\prod_{z=1}^{k}\Big(\min_{\lambda_{z-1}\in \Lambda}\sum_{\bar{\omega}_{\hat{x}k}(z)\in \hat{X}'_{-\epsilon}(\bar{\omega}_{qk}(z-1))}\!\!\!\!\!\!\!\!\!\!g(\bar{\omega}_{\hat{x}k}(z-1),\bar{\omega}_{qk}(z-1)),\hat{w}(z-1)\Big)r(\bar{\omega}_{\hat{x}k}(k),\bar{\omega}_{qk}(k),\hat{w}(k))\geq 1-\eta.
\end{equation*}
with $\hat{w}(m):=\lambda_m(\hat{x}(m),\rho'_{m}(\hat{x}(m),q(m)))$, $m\in[0,k-1]$.
Therefore, we have $\ul{V}^{\rho',\lambda^*(\rho')}_{H}(\hat{x}_0,\bar{q}_0) \leq \eta$ when applying the supervisor as in Definition~\ref{def:History-based Supervisor_wcv}.	 
According to\cite[Theorem 5.10]{Zhong2021Automata}, one has $\mathbb{P}_{\Omega}\Big\{y_{\omega H}\models \mathcal{A}\Big\}\leq \eta$, which completes the proof.	

\end{document}